\documentclass[12pt]{article}
\usepackage{eurosym}
\usepackage{amssymb}
\usepackage{epsfig}
\usepackage{graphicx}
\usepackage{amsmath}
\usepackage{amsfonts}
\usepackage{amssymb}
\usepackage{float}
\usepackage{subfigure}
\usepackage{float}
\usepackage{amsmath}
\usepackage{amsfonts}
\usepackage{amssymb}

\DeclareGraphicsExtensions{.pdf}
\graphicspath{{Images/}}
\begin{document}

\date{}
\title{\textbf{\Large Conductivity of higher dimensional holographic
superconductors with nonlinear electrodynamics}}
\author{ \textbf{{\normalsize Ahmad Sheykhi}$^{1,2} $\thanks{%
asheykhi@shirazu.ac.ir}}, \textbf{{\normalsize Doa Hashemi
Asl}$^{1}$},\,
\textbf{{\normalsize Amin Dehyadegari}$^{1}$} \\
$^{1}$ {\normalsize Physics Department and Biruni Observatory,
College of
Sciences,}\\
{\normalsize Shiraz University, Shiraz 71454, Iran}\\
$^{2}$ {\normalsize Research Institute for Astronomy and
Astrophysics of
Maragha (RIAAM),}\\
{\normalsize P.O. Box 55134-441, Maragha, Iran}}
\date{}
\maketitle

\begin{abstract}
We investigate analytically as well as numerically the properties
of $s$-wave holographic superconductors in $d$-dimensional
spacetime and in the presence of Logarithmic nonlinear
electrodynamics.  We study three aspects of this kind of
superconductors. First, we obtain, by employing analytical
Sturm-Liouville method as well as numerical shooting method, the
relation between critical temperature and charge density, $\rho$,
and disclose the effects of both nonlinear parameter $b$ and the
dimensions of spacetime, $d$, on the critical temperature $T_c$.
We find that in each dimension, $T_c/{\rho}^{1/(d-2)}$ decreases
with increasing the nonlinear parameter $b$ while it increases
with increasing the dimension of spacetime for a fixed value of
$b$. Then, we calculate the condensation value and critical
exponent of the system analytically and numerically and observe
that in each dimension, the dimensionless condensation get larger with
increasing the nonlinear parameter $b$. Besides, for a fixed value
of $b$, it increases with increasing the spacetime dimension. We
confirm that the results obtained from our analytical method  are
in agreement with the results obtained from numerical shooting
method. This fact further supports the correctness of our
analytical method. Finally, we explore the holographic
conductivity of this system and find out that the superconducting
gap increases with increasing either the nonlinear parameter or
the spacetime dimension.
\end{abstract}

\newpage \vspace*{0.2cm}
\section{Introduction}
One of the most important challenges, in the past decades, in
condensed matter physics is finding a justification for the high
temperature superconductors. The well-known
Bardeen-Cooper-Schrieffer (BCS) theory is the first successful
microscopic theory of superconductivity. This theory describes
superconductivity as a microscopic effect caused by a condensation
of Cooper pairs into a boson-like state \cite{BCS}. However, the
BCS theory is unable to explain the mechanism of the high
temperature superconductors in condensed matter physics. The
gauge/gravity duality or Anti de-Sitter (AdS)/Conformal Field
Theory (CFT) correspondence is a powerful tool which provides a
powerful tool for calculating correlation functions in a strongly
interacting field theory using a dual classical gravity
description \cite{Maldacena}. According to AdS/CFT correspondence,
the gravity theory in a $(d+1)$-dimensional AdS spacetime can be
related to a strong coupling conformal field theory on the
$d$-dimensional boundary of the spacetime.  The application of
this duality to condensed matter physics was suggested by Hartnoll
et.al., \cite{Har1,Har2} who suggested that some properties of
strongly coupled superconductors on the horizon of Schwarzschild
AdS black holes can be potentially described by the gravity theory
in the bulk, known as \textit{holographic superconductor}.
According to this proposal, a charged scalar field coupled to a
Maxwell gauge field is required in the black hole background to
form a scalar hair below the critical temperature. It was argued
that the coupling of the Abelian Higgs model to gravity in the
background of AdS spaces leads to black holes which spontaneously
break the gauge invariance via a charged scalar condensate
slightly outside their horizon \cite{Gub}. This corresponds to a
phase transition from black hole with no hair (normal
phase/conductor phase) to the case with scalar hair at low
temperatures (superconducting phase) \cite{Gub,CH,SAH}.

The properties of holographic superconductors have been
investigated extensively in the literature. When the gauge field
is in the form of linear Maxwell electrodynamics coupled to the
scalar field, the holographic superconductor has been explored in
\cite{Gary T.H,Gary T.H2, Wang4}. The studies on the holographic
superconductors have got a lot of attentions \cite{P.ZGJZ,
RGC2,Wang1,Wang2, RGC3, Ruth, GBHSC, Gr,
RGC5,RGC6,RGC7,XHWang,Wang3,Wang5,Wang6}. The investigation was
also generalized to nonlinear gauge fields such as Born-Infeld,
Exponential, Logarithmic  and Power-Maxwell electrodynamics.
Applying the analytical method including the Sturm-Liouville
eigenvalue problem \cite{SS,GR,SS2,SSM,GG,L.P.J.W,siopsis}, or
matching method which is based on the match of the solutions near
the boundary and on the horizon at some intermediate point
\cite{SSD,SS3}, or using the numerical method \cite{JC,JPC,ZPCJ},
the relation between critical temperature and charge density of
the $s$-wave holographic superconductors have been investigated.
It was argued that the nonlinear electrodynamics will affect the
formation of the scalar hair, the phase transition point, and the
gap frequency. In particular, with increasing the nonlinearity of
gauge field increases, the critical temperature of the
superconductor decreases and the condensation becomes harder,
however, it does not any affect on the critical exponent of the
system and it still obeys the mean field value
\cite{logholo,expholo,powerholo}.

In this paper, we explore the properties of the $s$-wave
holographic superconductor in higher dimension with Logarithmic
gauge field, by applying both the analytical Sturm-Liouville
eigenvalue method as well as numerical shooting method. In
particular, we disclose the effect of nonlinear electrodynamics
and the dimensions of the spacetime on the critical temperature of
the superconductor and its condensation. Also, we explore the
effects of nonlinearity as well as spacetime dimension on the gap
of frequency and electrical conductivity of the system. We shall
find that increasing the nonlinear parameter makes the
condensation harder, so the critical temperature decreases. In
addition, the gap frequency $\omega _{g}$ increases by increasing
the nonlinear parameter in each spacetime dimension.

This paper is outlined as follow. In the next section, we introduce
the basic field equations of the $d$-dimensional holographic
superconductor with Logarithmic nonlinear electrodynamics. In
section \ref{Crit}, we employ the Sturm-Liouville analytical
method as well as numerical shooting method to obtain a relation
between the critical temperature and charge density. We also
confirm that our analytical results are in agreement with the
numerical results. In section \ref{Exp}, we calculate analytically
and numerically, the critical exponent and the condensation value
of the system. In section \ref{Cond} we study the holographic
electrical conductivity of the system and reveal the response of
the system to an external field. The last section is devoted to
closing remarks.
\section{HSC with logarithmic nonlinear
electrodynamics in higher dimensions} \label{basic}

Our starting point is the $d$-dimensional action in the background
of AdS spacetime which includes Einstein gravity, nonlinear gauge
field, a scalar field and is described by
\begin{equation}\label{action}
S=\int d^{d}x\sqrt{-g}\left[ R-2\Lambda +\mathcal{L}_{m}\right] ,
\end{equation}%
where $R$ is the Ricci scalar, and
\begin{equation}
\Lambda =-\frac{(d-1)(d-2)}{2l^{2}},
\end{equation}
is the negative cosmological constant \cite{Alfonso V.R}, and $l$
is the AdS radius. The term $\mathcal{L}_{m}$ represents the
Lagrangian of the matter field, which is written as
\begin{equation*}
\mathcal{L}_{m}=-\frac{1}{b}\ln \bigg(1+\frac{b\mathcal{F}}{4}\Bigg)-|\nabla
\psi -iqA\psi |^{2}-m^{2}|\psi |^{2},
\end{equation*}%
where $ \mathcal{F}=F_{ab }F^{ab }$, $F_{ab }=\partial _{a }A_{b
}-\partial _{b }A_{a }$ is the electromagnetic field tensor, and
$A_{a}$ is the gauge field. The first term in the above expression
is the logarithmic Lagrangian which was introduced in
\cite{Soleng}, for the purpose of solving various divergencies in
the Maxwell theory. Here $b$ is the nonlinear parameter which
describes the strength of the nonlinearity of the theory. When
$b\rightarrow
0$, the logarithmic Lagrangian will reduce to the Maxwell form $\mathcal{L}=-\frac{1}{%
4}\mathcal{F}$. Also, $\psi $ is the scalar field with charge $q$ and mass
$m$.

Varying the action (\ref{action}) with respect to the metric
$g_{a b}$, the gauge field $A_a$ and the scalar field $\psi$
yields the following field equations
\begin{eqnarray}\label{enequation}
&&R_{ac}-\frac{g_{ac} R}{2}+\Lambda g_{ac}=-\frac{1}{2b}g_{ac}
\ln\left(1+\frac{b \mathcal{F}}{4}\right)
+\frac{2}{4+b \mathcal{F}} F_{ad} F_{c}^{\ d}-\frac{g_{ac}}{2}~m^2 |\psi|^2 \nonumber\\
 &&-\frac{g_{ac}}{2}
|\nabla\psi-iqA\psi|^2+\frac{1}{2}\left[(\nabla_{a}\psi-iqA_{a}\psi)(\nabla_{c}\psi^{*}+iqA_{c}\psi^{*})+a\leftrightarrow
c\right]~,
\end{eqnarray}
\begin{eqnarray}\label{gaugefield}
\nabla^{a}\left(\frac{4F_{ac}}{4+b
\mathcal{F}}\right)=iq\left[\psi^{*}(\nabla_{c}-iqA_{c})\psi-\psi(\nabla_{c}+iqA_{c})\psi^{*}\right]~,
\end{eqnarray}
\begin{eqnarray}\label{scalarfield}
(\nabla_{a}-iqA_{a})(\nabla^{a}-iqA^{a})\psi-m^2\psi=0.
\end{eqnarray}
When $b\rightarrow0$, the above equations reduce to the equations
of motion of holographic superconductors in Maxwell theory
\cite{Har2}. We shall work in the probe limit, in which the scalar
and gauge field do not back react on the metric background. We
consider a $d$-dimensional planar AdS-Schwarzschild black hole
which is described by the following metric
\begin{equation}\label{metric}
ds^2=-f(r)dt^2+\frac{1}{f(r)}dr^2+ r^2 h_{ij} dx^{i} dx^{j},
\end{equation}
where $h_{ij} dx^{i} dx^{j}$ is the line element of a $(d-2)
$-dimensional planar hypersurface and $f(r)$ is given by
\begin{eqnarray}\label{fequation}
f(r)=\frac{r^{2}}{l^{2}}-\frac{1}{r^{d-3}}\left(\frac{r^{d-1}_{+}}{l^{2}}%
\right),
\end{eqnarray}
where $r_{+}$ is  the event horizon radius. The temperature of the
superconductor is an important parameter in condensed matter
physics, so according to AdS/CFT dictionary, we need to have this
concept on the gravity side. The Hawking temperature of the black
hole on the horizon is given by
\begin{eqnarray} \label{T}
T=\frac{f^{\prime}(r_{+})}{4\pi}=\frac{(d-1)r_{+}}{4\pi l^{2}},
\end{eqnarray}
which should be identified as the temperature of the
superconductor. Here, the prime denotes derivative with respect to
$r$. Without lose of generality, we consider the electromagnetic
field and the scalar field in the forms
\begin{eqnarray}\label{Aandpsi}
A_{a}dx^{a}=\phi(r)~dt, \  \   ~~\psi=\psi(r).
\end{eqnarray}
Let us note that due to the gauge freedom, we can choose $\psi(r)$ to
be a real scalar field. Inserting metric (\ref{metric}) and scalar
and gauge fields (\ref{Aandpsi}) in the field equations
(\ref{gaugefield}) and (\ref{scalarfield}), we arrive at the
following equations for the gauge and scalar fields
\begin{eqnarray}  \label{phir}
\bigg(2+ b\phi^{\prime 2}(r)\bigg)\phi^{\prime\prime}(r)
+\frac{d-2}{r} \bigg(2- b \phi^{\prime 2}(r)\bigg)\phi^{\prime
}(r) -\frac{q^{2}\phi(r)}{ f(r)} \psi^{2} (r) \bigg(2- b
\phi^{\prime 2}(r)\bigg)^{2}=0,  \label{Eqphi}
\end{eqnarray}
\begin{eqnarray}  \label{psir}
\psi^{\prime\prime}(r) +\left(\frac{f^{\prime
}}{f}+\frac{d-2}{r}\right) \psi^{\prime }(r) +\bigg(\frac
{q^{2}\phi^2(r)}{f^2(r)}-\frac{m^2}{f(r)}\bigg) \psi(r)=0.
\end{eqnarray}
Our next step is to solve the nonlinear field equations
(\ref{phir}) and  (\ref{psir}) and obtain the behavior of
$\psi(r)$ and $\phi(r)$. For this purpose we need to fix the
boundary conditions for $\phi(r)$ and $\psi(r)$ at the black hole horizon ($%
r=r_+$) and at the asymptotic AdS boundary ($r\rightarrow
\infty$). From Eqs.(\ref{phir}) and (\ref{psir}), and using the
fact that $f(r_{+})=0$, we can imply the boundary conditions
\begin{eqnarray}\label{boundrplus}
\phi(r_{+})=0,~~~
\psi(r_{+})=\frac{f^\prime(r_{+})\psi^\prime(r_{+})}{m^{2}}.
\end{eqnarray}
The reason that $A_{t}=\phi(r)$, must be zero at the horizon comes
from the fact that at the horizon, the quantity $A^2=
g^{a b}A_{a}A_{b}$, that is the norm of vector, should be
finite at $r=r_{+}$. Far from the horizon boundary, at the spatial
infinity ($r\rightarrow\infty$), the asymptotic performance of the
solutions are
\begin{eqnarray}  \label{bound1}
\phi(r)=\mu-\frac{\rho}{r^{d-3}},
\end{eqnarray}
\begin{eqnarray}  \label{bound2}
\psi(r)=\frac{\psi_{-}}{r^{\Delta_{-}}}+\frac{\psi_{+}}{r^{\Delta_{+}}},
\end{eqnarray}
where
\begin{eqnarray}\label{delta}
\Delta_{\pm}=\frac{1}{2}\left[(d-1)\pm\sqrt{(d-1)^2+4m^2
l^2}\right].
\end{eqnarray}
Here the parameters $\mu$ and $\rho$ are dual to chemical
potential and charge density of the field theory on the boundary.
Coupling the scalar field to the Maxwell field in the field
equations give us an effective mass for $\psi$ that can be
positive or negative, but since, at low temperature it is possible
that the effective mass becomes sufficiently negative, so in this
temperature we have an instability in the formation of the scalar
field and the system will encounter the superconducting phase
\cite{Gary T.H2}. Thus, we can have negative mass for $\psi$ but
it must satisfy the BF (Breitenlohner-Freedman) bound
\cite{Alfonso V.R},
\begin{eqnarray}\label{BF}
m^{2}\geq -\frac{(d-1)^{2}}{4l^{2}},
\end{eqnarray}
which can be easily understood from Eq. (\ref{delta}). In what
follows we will choose some values for $m^2$ that satisfy this
bound.
From the AdS/CFT dictionary, we have $%
\langle\mathcal{O}\rangle$ as a condensation operator on the
boundary, which is dual to the scalar field $\psi$ in the bulk. We
can choose the boundary condition in which either $\psi_{+}$ or
$\psi_{-}$ vanishes \cite{Har1}. Indeed, either $\psi_{+}$ or
$\psi_{-}$ can be dual to the value of the operator, and the other
one is dual to its source. However to keep up the stability of
the AdS space, one of them must be equal to zero \cite{Gary T.H}.
In this paper we shall choose $\psi_{-}=0$ and take
$\psi_{+}=\langle\mathcal{O_{+}}\rangle $ non zero.

It is worth noting that Eqs.(\ref{phir}) and  (\ref{psir}) have
several scaling symmetries, one of them is,
\begin{eqnarray}\label{scalingsym}
\phi\rightarrow a\phi,~~\psi\rightarrow a\psi,~~q\rightarrow
qa^{-1},~~b\rightarrow b a^{-2}~,
\end{eqnarray}
This symmetry allows us to choose $q=1$ in the equations,
without loss of generality. We shall also choose $l=1$ by using other symmetries. In the
remaining part of this paper, we study analytically as well as
numerically the different properties of the HSC with nonlinear
electrodynamics.
\section{Critical temperature versus charge density} \label{Crit}
In this section, we would like to explore the critical temperature
of higher dimensional HSC in the presence of logarithmic nonlinear
electrodynamics. Our investigation will be both analytically and
numerically. At the end of this section, we compare our results.
\subsection{Analytical method}
First, we obtain  analytically a relation between the critical
temperature and charge density of the HSC by using the
Sturm-Liouville eigenvalue problem. For convenience, we transform
the coordinate in such a way that, $r\rightarrow z={r_{+}}/{r}$.
Under this transformation, Eqs.(\ref{phir}) and (\ref{psir}) can
be rewritten as
\begin{eqnarray}\label{phiz}
&&\left(2+b\frac{z^4}{r_{+}^2}\phi^{\prime 2}\right)\phi^{\prime%
\prime}+\frac{b z^{3}d }{r_{+} ^{2}} {{\phi^{\prime}}^{3}}-\frac{
r_{+}^2}{z^4}\frac{\phi}{f}\psi^2\bigg(2- b \frac{z^4}{r_{+}^2}%
\phi^{\prime 2}\bigg)^2 +\frac{2(4-d)}{z}{\phi^{\prime }}=0,
\end{eqnarray}
\begin{eqnarray}\label{psiz}
\psi^{\prime\prime}+\bigg(\frac{f^{\prime}}{f}+\frac{4-d}{z} \bigg) %
\psi^{\prime}+\bigg(\frac{%
r_{+}^2\phi^2}{z^4 f^2}-\frac{m^{2} r_{+}^{2}}{z^4 f}
\bigg)\psi=0.
\end{eqnarray}
where the prime now indicates derivative with respect to $z$. At
the critical temperature ($T=T_c$) we have $\psi=0$, which implies
that in this temperature the condensation is zero. Thus, Eq.(\ref{phiz}) reduces to
\begin{equation}\label{pphii}
\bigg(2+ b \frac{z^4}{r_{+c}^2}\phi^{\prime 2} \bigg)\phi^{\prime%
\prime}(z) + \frac{b z^{3}d }{r_{+c}^{2}} {{\phi^{\prime}}^{3}(z)} + \frac{%
2(4-d)}{z}\phi^{\prime }(z)=0~,
\end{equation}
Now, we try to solve the above equation and find a solution for
this equation in the interval $[z,1]$. Considering the asymptotic
behavior of $\phi$ near the AdS boundary ($z\rightarrow0$), given
in Eq.(\ref{bound1}), we can write the solution in the form
\begin{eqnarray}\label{phi0}
\phi(z)=\lambda r_{+_{c}} \zeta(z),
\end{eqnarray}
where $\lambda=\frac{\rho}{r_{+c}^{d-2}}$, and
\begin{eqnarray}\label{zeta}
\zeta(z)=\int^{1}_{z}\frac{\sqrt{1+2(d-3)^2 b\lambda^2{\tilde{z}}^{(2d-4)}}-1}{(d-3)b\lambda^2 \tilde{z}^d}d\tilde{z},
\end{eqnarray}
and we have used the fact that $\phi(1)=0$. Since the above
integral cannot be solved exactly, we perform a perturbative
expansion of $2(d-3)^2 b\lambda^{2}$ in the right side of
Eq.(\ref{zeta}) and consider only the terms that are linear in
$b$. For this purpose, we assume the nonlinear parameter $b$
expressed as
\begin{eqnarray}  \label{BIbn}
b_{n}=n\Delta b,\  \   \   \  \   ~~~n=0,1,2,\cdot\cdot\cdot~,
\end{eqnarray}
when $\Delta b=b_{n+1}-b_{n}$ \cite{L.P.J.W}. So the expansion of $2(d-3)^2
b\lambda^{2}$ is
\begin{eqnarray}  \label{bLambda}
2(d-3)^{2}b\lambda^{2}=2(d-3)^{2}b_{n}\lambda^{2}=2(d-3)^{2}b_{n}(%
\lambda^{2}|_{b_{n-1}})+O[(\Delta b)^{2}]~.
\end{eqnarray}
Substituting Eq.(\ref{bLambda}) into Eq.(\ref{zeta}), we can distinguish two cases \cite{L.P.J.W}:%
\newline
In the first case where
$2(d-3)^{2}b_{n}(\lambda^{2}|_{b_{n-1}})<1$, we have
\begin{eqnarray}\label{zeta1}
&\zeta(z)&=\zeta_{1}(z)\approx\int^{1}_{z}\frac{[1+b_{n}(\lambda^2|_{b_{n-1}})(d-3)^2 {\tilde{z}}^{(2d-4)}-\frac{1}{2}~{b_{n}}^2 (\lambda^4|_{b_{n-1}}) (d-3)^4{\tilde{z}}^{(4d-8)}+...]-1}{(d-3)b_{n}(\lambda^2|_{b_{n-1}})\tilde{z}^{d}} d\tilde{z}
\notag \\
&&=(1-z^{d-3})+\frac{(d-3)^{3}b_{n}(\lambda^{2}|_{b_{n-1}})}{2(7-3d)}%
(1-z^{3d-7}).  \label{ZetaCase1}
\end{eqnarray}
In the second case where
$2(d-3)^{2}b_{n}(\lambda^{2}|_{b_{n-1}})>1$, the integration can be
done for two ranges of values of $z$, one for $z<z_{0}<1$ and
the other for $z_{0}< z\leq1$. Here $z_{0}$ is obtained from
$2(d-3)^{2}b_{n}(\lambda^{2}|_{b_{n-1}})z^{2(d-2)}=1$ for
$z=z_{0}$. In the former case where $z<z_{0}<1$, we have,

\begin{eqnarray}\label{zeta2}
&\zeta(z)&=\zeta_{2}(z)\approx\int^{z_{0}}_{z}\frac{[1+b_{n}(\lambda^2|_{b_{n-1}})(d-3)^2 {\tilde{z}}^{(2d-4)}-\frac{1}{2}~{b_{n}}^2 (\lambda^4|_{b_{n-1}}) (d-3)^4{\tilde{z}}^{(4d-8)}+...]-1}{(d-3)b_{n}(\lambda^2|_{b_{n-1}})\tilde{z}^{d}} d\tilde{z}  \notag \\
&&+\int^{1}_{z_{0}} \frac{1}{(d-3)b_{n} (\lambda|_{b_{n-1}}) {\tilde{z}}^{d}}\bigg[\sqrt{2b_{n}}(\lambda|_{b_{n-1}})(d-3)\tilde{z}^{(d-2)}\bigg(1+\frac{1}{4b_{n}(\lambda^2|_{b_{n-1}})(d-3)^2\tilde{z}^{(2d-4)}}   \notag \\
&&-\frac{1}{32~b_{n}^2 (\lambda^4|_{b_{n-1}})(d-3)^4 \tilde{z}^{(4d-8)}}\bigg)-1\bigg]d\tilde{z}  \notag \\
&&=-z^{d-3}+z_{0}^{d-3}-\frac{\left(d-3\right)^3}{2\left(3d-7\right)}%
b_{n}\left(\lambda^{2}|_{b_{n-1}}\right)\left(z_{0}^{3d-7}-z^{3d-7}\right)+%
\frac{\sqrt{2}}{\sqrt{b_{n}}(\lambda|_{b_{n-1}})}\left(\frac{1}{z_{0}}%
-1\right)  \notag \\
&&+\frac{1}{2\sqrt{2}\left(-2d+3\right)\left(d-3\right)^2 b_{n}\sqrt{b_{n}}%
    \left(\lambda^{3}|_{b_{n-1}}\right)}\left(1-z_{0}^{-2d+3}\right)-\frac{1}{(d-3)(-d+1)b_{n}(\lambda^2|_{b_{n-1}})} \notag \\
&&\left(1-z_{0}^{-d+1}\right)-\frac{1}{16\sqrt{2}(d-3)^4(-4d+7)~b_{n}^2~\sqrt{b_{n}}(\lambda^5|_{b_{n-1}})}\left(1-z_{0}^{-4d+7}\right).
\end{eqnarray}
Since we have
\begin{eqnarray}\label{blambda}
b_{n}(\lambda^2|_{b_{n-1}})=\frac{1}{2(d-3)^2{z_{0}}^{2d-4}},
\end{eqnarray}
thus Eq.(\ref{zeta2}) can be written in terms of $z_{0}$,
\begin{eqnarray}\label{zeta21}
&&\zeta_{2}(z)=-z^{d-3}+z_{0}^{d-3}-\frac{(d-3)}{4(3d-7)z_{0}^{2d-4}}\left(z_{0}^{3d-7}-z^{3d-7}\right)+%
2(d-3)z_{0}^{d-2}\left(\frac{1}{z_{0}}%
-1\right) \notag \\
&&+\frac{(d-3)z_{0}^{3d-6}}{-2d+3}\left(1-z_{0}^{-2d+3}\right)-\frac{2(d-3)z_{0}^{2d-4}}{(-d+1)}\left(1-z_{0}^{-d+1}\right) \notag \\
&&-\frac{(d-3)z_{0}^{5d-10}}{4(-4d+7)}\left(1-z_{0}^{-4d+7}\right).
\end{eqnarray}
While in the latter case where $z_{0}< z\leq1$, we find
\begin{eqnarray}\label{zeta3}
\zeta(z)&=&\zeta_{3}(z)\approx\int^{1}_{z} \frac{1}{(d-3)b_{n} (\lambda|_{b_{n-1}}) {\tilde{z}}^{d}}\bigg[\sqrt{2b_{n}}(\lambda|_{b_{n-1}})(d-3)\tilde{z}^{(d-2)}\bigg(1+\frac{1}{4b_{n}(\lambda^2|_{b_{n-1}})(d-3)^2\tilde{z}^{(2d-4)}}   \notag \\
&&-\frac{1}{32~b_{n}^2 (\lambda^4|_{b_{n-1}})(d-3)^4 \tilde{z}^{(4d-8)}}\bigg)-1\bigg]d\tilde{z} \notag \\
&&=\frac{\sqrt{2}}{\sqrt{b_{n}}(\lambda|_{b_{n-1}})}\left(\frac{1}{z}%
-1\right) +\frac{1}{2\sqrt{2}\left(-2d+3\right)\left(d-3\right)^2 b_{n}\sqrt{b_{n}}%
    \left(\lambda^{3}|_{b_{n-1}}\right)}\left(1-z^{-2d+3}\right) \notag \\
&&-\frac{1}{(d-3)(-d+1)b_{n}(\lambda^2|_{b_{n-1}})}\left(1-z^{-d+1}\right) \notag \\
&&-\frac{1}{16\sqrt{2}(d-3)^4(-4d+7)~b_{n}^2~\sqrt{b_{n}}(\lambda^5|_{b_{n-1}})}\left(1-z^{-4d+7}\right).
\end{eqnarray}

and from Eq.(\ref{blambda}),
\begin{eqnarray}\label{zeta31}
&&\zeta_{3}(z)=2(d-3)z_{0}^{d-2}\left(\frac{1}{z}-1\right) +\frac{(d-3)z_{0}^{3d-6}}{-2d+3}\left(1-z^{-2d+3}\right) \notag \\
&&-\frac{2(d-3)z_{0}^{2d-4}}{-d+1}\left(1-z^{-d+1}\right)-\frac{(d-3)z_{0}^{5d-10}}{4(-4d+7)}\left(1-z^{-4d+7}\right).
\end{eqnarray}
At the first approximation the asymptotic AdS boundary condition
for $\psi$ is given by Eq.(\ref{bound2}). Near the asymptotic AdS
boundary, we define a function $F(z)$ such that
\begin{eqnarray} \label{psiF}
\psi(z)\sim\frac{\langle\mathcal{O_{+}}\rangle}{r_{+}^{\Delta_{+}}}
z^{\Delta_{+}}F(z).
\end{eqnarray}
Substituting Eq.(\ref{psiF}) into Eq.(\ref{psiz}), we arrive at
\begin{eqnarray}\label{Fequation}
&&F^{\prime\prime}(z)+F^{\prime }(z)\left(\frac{4-d+2\Delta_{+}}{z}+\frac{%
f^{\prime}(z)}{f(z)}\right)+F(z)\left(\frac{\Delta_{+}(3-d+\Delta_{+})}{z^{2}}+\frac{%
\Delta_{+} f^{\prime}(z)}{z f(z)}-\frac{m^{2}r_{+}^{2}}{z^4
f(z)}\right)\nonumber\\
&&+ F(z)\left(\frac{r_{+c}^{4}\lambda^{2}\zeta^{2}(z)}{z^4 f^{2}(z)}%
\right)=0.
\end{eqnarray}
The above equation can be written in the Sturm-Liouville form
\begin{equation}\label{sturm}
[T(z)F^{\prime}(z)]^{\prime}-Q(z)F(z)+\lambda^{2}F(z)N(z)\zeta^2(z)=0~,
\end{equation}
where we have defined
\begin{eqnarray}
T(z)&=&r_{+c}^{2} z^{2\Delta_{+}+2-d}(1-z^{d-1}),\nonumber
\\
Q(z)&=&r_{+c}^{2} z^{2\Delta_{+}-d-1}\big[ z^{d} \Delta_{+}^{2}+z
(m^{2}-\Delta_{+}+\Delta_{+} d-\Delta_{+}^{2})\big],
\\
N(z)&=&\frac{1}{(1-z^{d-1})^{2}}.\nonumber
\end{eqnarray}
According to the Sturm-Liouville eigenvalue problem, the eigenvalues of Eq.
(\ref{sturm}) are
\begin{eqnarray}  \label{lambdaeigenvalueCase1}
\lambda^{2}=\frac{\int^{1}_{0}(TF^{\prime 2}+QF^{2})dz}{\int^{1}_{0}TN%
\zeta_{1}^{2}F^2dz}\ , & \quad   \    \  \  \  \mathrm{for}~ \
2(d-3)^{2}b_{n}(\lambda^{2}|_{b_{n-1}})<1,
\end{eqnarray}
and
\begin{eqnarray}  \label{lambdaeigenvalueCase2}
\lambda^{2}=\frac{\int^{1}_{0}(TF^{\prime 2}+QF^{2})dz} {\int^{%
z_{0}}_{0}TN\zeta_{2}^{2}F^2dz+\int^{1}_{z_{0}}TN\zeta_{3}^{2}F^2dz}\
, & \quad \mathrm{for}~ \ 2(d-3)^{2}b_{n}(\lambda^{2}|_{b_{n-1}})>1.
\end{eqnarray}
 we assume  the trial function
$F(z)$ in the form\cite{siopsis},
\begin{equation}
F(z)=1-\alpha z^{2},
\end{equation}
which satisfies the boundary conditions $F(0)=1$ and $F^{\prime
}(0)=0$.
 We now can determine $\lambda^{2}$ for different values of parameters $d$ and $b$. From Eq.(\ref{T}) at the
critical point, the temperature is
\begin{eqnarray}\label{TTc}
T_{c}=\frac{(d-1) r_{+c}}{4\pi}.
\end{eqnarray}
Using the fact that $\lambda={\rho}/{r_{+c}^{d-2}}$, we can
rewrite the critical temperature for condensation in terms of the
charge density $\rho$ as
\begin{eqnarray}\label{Tc}
T_{c}=\frac{(d-1)}{4\pi}\left(\frac{\rho}{\lambda}\right)^{\frac{1}{d-2}}.
\end{eqnarray}
This implies that the critical temperature is proportional to
$\rho ^{1/(d-2)}$. According to our analytical method, in order to
calculate the critical temperature for the condensation, we
minimize the function $\lambda$ in
Eqs.(\ref{lambdaeigenvalueCase1}) and
(\ref{lambdaeigenvalueCase2}) with respect to the coefficient
$\alpha$ for different values of nonlinear parameter $b$ and
spacetime dimension $d$. Then, we obtain ${T_{c}}/{[
\rho^{1/(d-2)}]}$ through relation (\ref{Tc}). As an example, we
bring the details of our  calculation for $d=5$, $n=1$ and the
step size  $\Delta b=0.1$. From Eq. (\ref{BIbn}), we have
$b_{1}=0.1$. At first, we must calculate
$2(d-3)^{2}b_{n}(\lambda^{2}|_{b_{n-1}})$ for this case, to find
out which equation for obtaining $\lambda^2$ should be used. We
find $\lambda^{2}|_{b_{n-1}}=\lambda^{2}|_{b_{0}=0}=18.22$. Thus,
$2(d-3)^{2}b_{1}(\lambda^{2}|_{b_{0}})=14.58$. This indicates that
we should use Eq.(\ref{lambdaeigenvalueCase2}). This equation for
the fixed $d$ and $b$ reduces to
\begin{eqnarray}
\lambda^2=\frac{380.64-570.96\alpha+296.05\alpha^2}{4.73-3.84\alpha+\alpha^2},
\end{eqnarray}
which its minimum is $\lambda_{\mathrm{min}}^2=49.25$ for
$\alpha=0.773$. We use this value for calculate the critical
temperature. The critical temperature becomes
$T_{c}=0.166\rho^{1/3}$. In tables (\ref{t1}), (\ref{t2}) and
(\ref{t3}), we summarize our results for $\lambda _{\mathrm{min}}$
and ${T_{c}}/{[ \rho^{1/(d-2)}]}$ for different values of the
parameters $d$ and $b$. From these tables we see that at a fixed
$d$, the critical temperature decrease as the nonlinear parameter
$b$ increases and for a fixed value $d$ the critical temperature
increase by increasing $d$.
\subsection{Numerical method}
In this subsection we study numerically the critical behavior of
the logarithmic holographic superconductor. For this purpose we
use the shooting method. We have the second-order Eqs.(\ref{phir}) and (\ref{psir}). For solving these equations, we
require four initial values on the horizon, namely $\phi(r_{+})$,
$\phi^{^{\prime }}(r_{+})$, $\psi(r_{+})$ and $\psi^{^{\prime
}}(r_{+})$. But with regards to Eq.(\ref{boundrplus}),
$\psi^{^{\prime }}(r_{+})$ and $\psi(r_{+})$ are not independent,
also $\phi(r_{+})=0$. So we just have two parameter at the horizon
that are
independent, they are $\psi(r_{+})$ and $\phi^{^{\prime }}(r_{+})$. Note that $%
\phi^{^{\prime }}(r_{+})$ means the value of the electric field at
the horizon ($\phi^{\prime 2}=F_{a b} F^{a b}$).

Also these equations have two other scaling symmetries except
Eq.(\ref{scalingsym}), which
allow us to set $r_{+}=1$ and $l=1$ to perform numerical calculation \cite%
{Har1}. After using these scalings, only two parameters that
specify the initial values at the horizon ($\psi(r_{+})$ ,
$\phi^{\prime }(r_{+})$) are determinative for our numerical
calculation. Therefore, the $\phi$ and $\psi$ equations in $z$
coordinate becomes
\begin{eqnarray}\label{phiphi}
&&\bigg(2+b z^4\phi^{\prime 2}(z) \bigg)\phi^{\prime\prime}(z)+bdz^{3} {{%
\phi^{\prime}}^{3}(z)} -\frac{\phi(z)}{z^4 f(z)}\psi^2(z)\bigg(2-
b z^4\phi^{\prime 2}(z)\bigg)^2
\nonumber\\&&+\frac{(-2d+8)}{z}{\phi^{\prime }(z)}=0,
\end{eqnarray}
\begin{eqnarray}\label{psipsi}
\psi^{\prime\prime}(z)+\bigg(\frac{f^{\prime}(z)}{f(z)}+\frac{4-d}{z} \bigg) %
\psi^{\prime}(z)+\bigg(-\frac{m^{2}}{z^4 f(z)}+\frac{\phi^2(z)}{z^4 f^2(z)} %
\bigg)\psi(z)=0.
\end{eqnarray}
To obtain initial values, we consider the behavior of the $%
\psi $ and $\phi$ near the horizon $(z=1)$, such that
\begin{eqnarray}\label{psiexpantion}
\psi\approx\psi(1)-\psi^{\prime }(1)(1-z)+\frac{\psi^{\prime \prime }(1)}{2}%
(1-z)^2+...,
\end{eqnarray}
\begin{eqnarray}\label{phiexpantion}
\phi\approx-\phi^{\prime }(1)(1-z)+\frac{\phi^{\prime \prime }(1)}{2}%
(1-z)^2+....
\end{eqnarray}
{According to these expansions, we find that the
coefficients which are determinative for calculating $\phi$ and $\psi$, are in the form $%
\phi(1)$, $\phi^{\prime }(1)$, $\psi(1)$, $\psi^{\prime }(1)$,
$\phi^{\prime \prime}(1)$,  $\psi^{\prime \prime }(1)$ and ....
The effects of coefficients of $(1-z)^{n}$ when $n$ is large, can
be  neglected. Because the value of  $(1-z)^{n}$ in higher orders
are very small in the vicinity of the horizon where $z=1$. Also,
we set $\phi(1)=0$ in Eq. (\ref{phiexpantion}). If we substitute
these expansions into Eqs.(\ref{phiphi}) and (\ref{psipsi}), we
can find all these coefficients \big($\phi^{\prime }(1)$,
$\phi^{\prime \prime}(1)$,  $\phi^{\prime \prime \prime}(1)$,
$\phi^{(4)}(1)$, $\psi(1)$, $\psi^{\prime }(1)$, $\psi^{\prime
\prime }(1)$, $\psi^{\prime \prime \prime}(1)$, $\psi^{(4) }(1)$
\big) in terms of $\psi(1)$ and $\phi^{\prime }(1)$. Thus, as
before mentioned only the values $\psi(1)$ and $\phi^{\prime }(1)$
are determinative.} Near the critical temperature $\psi $ is very
small, so we can set $\psi(1)=0.00001$. According to the shooting
method we can perform numerical calculation near the horizon with
one shooting parameter $\phi^{\prime }(1)$, to get proper
solutions at infinite boundary. This value of $\phi^{\prime }(1)$
can give us the value of critical density $\rho_{c}$ through
Eq.(\ref{phi0}).

By solving equations numerically, we find that $\phi $ is a
uniform function that starts at zero value at the horizon and
increases to the value $\mu$ in the asymptotic boundary. But for
$\psi$, there are unlimited solutions that satisfy our
boundary condition. We can label this solutions by number of times that $%
\psi $ get zero in the interval $[0,1]$. From these solutions only
the case that reduces uniformly from $\psi(1)$ to zero, will be
stable. In \figurename{1} the various solutions for $d=4$ and
$d=5$ has been shown, where the blue line shows stable $\psi$
.\newline
\begin{figure}[H]
\centering
\subfigure[$d=4$, $b=0$] {\includegraphics[scale=0.85]{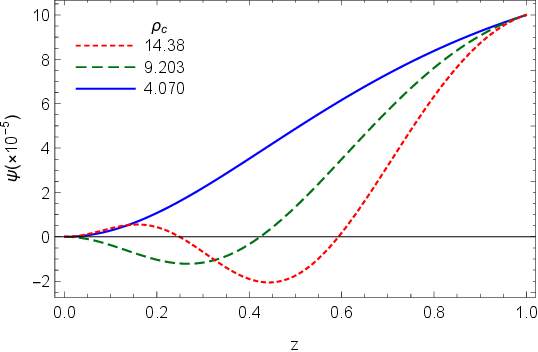}} \label%
{fig1a} \hspace*{.1cm}
\subfigure[$d=5$,
$b=0$]{\includegraphics[scale=0.85]{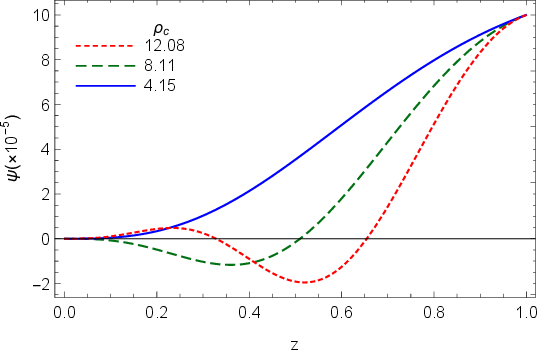}} \label{fig1b} \hspace*{.2cm}
\caption{$\protect\psi(z)$ with boundary condition $\protect\psi_{-}=0$ for
three lowest values of $\protect\rho_{c}$.}
\label{fig1}
\end{figure}
We can determine $\phi$ and $\psi$ by using the numerical
calculation. Thus we can find the coefficients in the asymptotic
behavior of these fields Eqs.(\ref{bound1}) and (\ref{bound2}),
which are $\mu,\rho$ and $\psi_{+}$(we choose $\psi_{-}$ to be
zero). By specifying the values of $\rho$, we can find the
resealed critical temperature. In tables (\ref{t1}-\ref{t3}), we
summarize the results for the critical temperature of phase
transition of holographic superconductor in the presence of
logarithmic nonlinear electrodynamics for different values of $b$
and $d$. Also we compare the analytical results obtained from
Sturm-Liouville method with those obtained in this subsection
numerically. From this table, we observe that the analytical
results are in good agreement with the numerical results.

\begin{figure}[H]
\begin{center}
\begin{minipage}[b]{0.325\textwidth}\begin{center}
        \subfigure[~$d=4,~b=0.2$]{
                 \label{fig2a}\includegraphics[width=\textwidth]{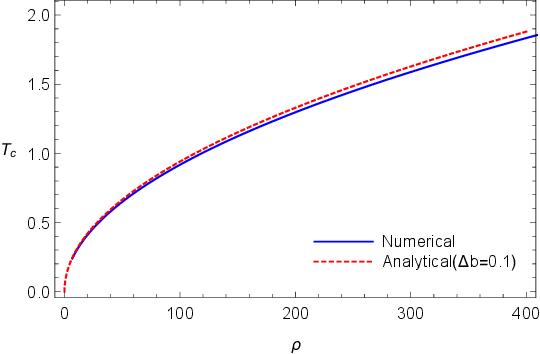}\qquad}
     \end{center}\end{minipage} \hskip+0cm
\begin{minipage}[b]{0.325\textwidth}\begin{center}
         \subfigure[~$d=5,~b=0.2$]{
                  \label{fig2b}\includegraphics[width=\textwidth]{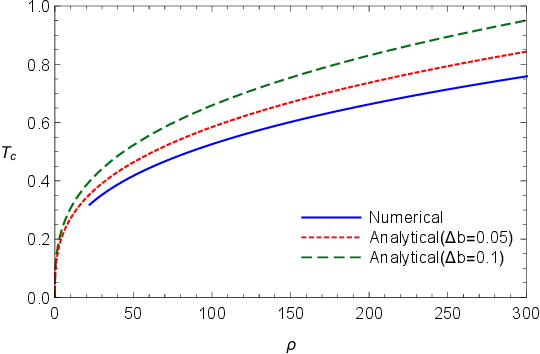}\qquad}
     \end{center}\end{minipage} \hskip0cm
\begin{minipage}[b]{0.325\textwidth}\begin{center}
          \subfigure[~$d=6,~b=0.1$]{
                   \label{fig2c}\includegraphics[width=\textwidth]{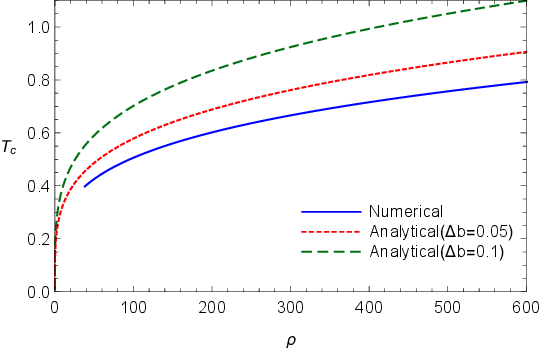}\qquad}
     \end{center}\end{minipage} \hskip0cm
\end{center}
\caption{Comparison of $T_{c}$ in terms of $\protect\rho$ from
numerical and analytical calculation.} \label{fig2}
\end{figure}

In Table (\ref{t1}) we show the critical temperature for different values
of $b$ with the scalar operator
$\langle\mathcal{O}\rangle=\langle\mathcal{O_{+}}\rangle$ for
3-dimensional superconductor, we consider $m^{2}=-$2 and $\Delta
b=0.1$. We see that the values obtained analytically with this
step size, is indeed in very good agreement with the numerical
results.
\begin{table}[tbh]
\label{table1} \centering
\begin{tabular}{ccccccc}
\hline\hline $b$   & $\alpha$   & $\lambda^2_{\rm min}$   &
$T_{c}\mid_{\rm Analytical}$   & $T_{c}\mid_{\rm Numerical}$   &  \\%
[0.05ex] \hline
$0$   &   0.601 &   17.30 &   0.117$\sqrt{\rho}$   & 0.118 $\sqrt{\rho}$ \\
$0.1$ &   0.632 &   27.79 &   0.103$\sqrt{\rho}$ &   0.103$\sqrt{\rho}$ \\
$0.2$ &   0.652 &   41.55 &   0.094$\sqrt{\rho}$ &   0.092$\sqrt{\rho}$ \\
$0.3$ &   0.668 &   61.16 &   0.085$\sqrt{\rho}$ &   0.082$\sqrt{\rho}$ \\[0.5ex] \hline
\end{tabular}%
\caption{Critical temperature $T_{c}$ for $d=4$ and
$\Delta_{+}=2$, $m^{2}=-2$. Here we choose the step size $\Delta
b=0.1$.}\label{t1}
\end{table}
Similarly in Table (\ref{t2}) and (\ref{t3}), we show the critical temperature for
different values of $b$, of the scalar operator for $4$ and $5$
dimensional superconductor with the mass of scalar field $m^2=-3$
and $m^2=-4$. We
choose the step size $\Delta b=0.1$ and $\Delta b=0.05$. For the case $%
\Delta b=0.05$, the agreement of analytical results derived from
Sturm-Liouville method with the numerical calculation is clearly
seen. So by reducing the step size in higher dimensions, we can
improve the analytical result.
\begin{table}[H]
\centering
\par
\begin{tabular}{ccccccccccccccc}
 \hline\hline
b &  & $\alpha_{1}$ &  & $\lambda^2_{\rm min(1)}$ &  &
$T_{c}|_{\rm Analytical(1)}$ &  & $\alpha_{2}$ &  &
$\lambda^2_{\rm min(2)}$ &  & $T_{c}|_{\rm Analytical(2)}$ &  &
$T_{c}|_{\rm Numerical}$ \\ \hline
$0$ &  & 0.722 & & 18.23 & & 0.196 $\rho^{1/3}$ & & 0.722 & &18.23 & &0.196 $\rho^{1/3}$ & & 0.198$\rho^{1/3}$ \\
$0.1$ &  & 0.773 &  & 49.25 & & 0.166 $\rho^{1/3}$ & & 0.786& & 70.38 & &0.156$\rho^{1/3}$ & & 0.145$\rho^{1/3}$ \\
$0.2$ &  & 0.804 & & 126.79 & & 0.142 $\rho^{1/3}$ & &0.820 & & 251.15 & &0.126 $\rho^{1/3}$& & 0.113$\rho^{1/3}$ \\
$0.3$ &  & 0.825 & & 327.92 & & 0.121  $\rho^{1/3}$& &0.841 & & 942.89 & &0.101$\rho^{1/3}$ & & 0.090$\rho^{1/3}$ \\  \hline
\end{tabular}%
\caption{ Critical temperature $T_{c}$ for $d=5$ and
$\Delta_{+}=3, m^2=-3$. Here the step size for Analytical(1) is
$\Delta b=0.1$ and for Analytical(2) is $\Delta b=0.05$. }\label{t2}
\end{table}
\begin{table}[H]
    \centering
    \par
    \begin{tabular}{ccccccccccccccc}
        \hline\hline
         b &  & $\alpha_{1}$ &  & $\lambda^2_{\rm min(1)}$ &  & $T_{c}|_{\rm Analytical(1)}$ &  & $\alpha_{2}$ &  & $\lambda^2_{\rm min(2)}$ &
        & $T_{c}|_{\rm Analytical(2)}$ &  & $T_{c}|_{\rm Numerical}$ \\ \hline
        $0$ &  & 0.792 & & 22.66 & & 0.269 $\rho^{1/4}$ & & 0.792 & &22.66 & &0.269 $\rho^{1/4}$ & & 0.271$\rho^{1/4}$ \\
        $0.1$ &  & 0.854 &  & 104.14 & & 0.222 $\rho^{1/4}$ & & 0.878& &494.17 & &0.183$\rho^{1/4}$ & & 0.160$\rho^{1/4}$ \\
        $0.2$ &  & 0.880 & & 471.19 & & 0.184 $\rho^{1/4}$ & &0.909 & & 19538.1 & &0.116 $\rho^{1/4}$& & 0.104$\rho^{1/4}$ \\
        $0.3$ &  & 0.902 & & 5283.9 & & 0.136  $\rho^{1/4}$& &0.920 & & 1.1$\times 10^{6}$ & &0.069$\rho^{1/4}$ & & 0.067$\rho^{1/4}$ \\  \hline
    \end{tabular}%
\caption{Critical temperature $T_{c}$ for $d=6$ and $\Delta_{+}=4,
m^2=-4$. Here the step size for Analytical(1) is $\Delta b=0.1$
and for Analytical(2) is $\Delta b=0.05$. }\label{t3}
\end{table}
{It is worth noting that according to the BF bound given in Eq.
(\ref{BF}), the mass of the scalar field, depends on the spacetime
dimension. For example, $m^2\geq -{9}/{4}$ for $d=4$, $m^2\geq -4$
for $d=5$, and $m^2\geq -{25}/{4}$ for $d=6$. For convenient, in
this paper we choose the mass as $m^2=-2,-3,-4$ for $d=4,5,6$,
respectively. The reason for these choice comes from the fact that
for these values of $m$, the value of $\Delta_{+}$ becomes integer
and so the calculations are simplified. In addition, if we assume
a fixed value for $m$ in all of these dimensions, we arrive at the
same result (see table $4$).}
\begin{table}[H]
    \label{table1} \centering
    \begin{tabular}{ccccccc}
        \hline\hline $d$   & $\alpha$   & $\lambda^2_{\rm min}$   &
        $T_{c}\mid_{\rm Analytical}$   & $T_{c}\mid_{\rm Numerical}$   &  \\%
        [0.05ex] \hline
        $4$   &   0.601 &   17.30 &   0.117$\sqrt{\rho}$   & 0.118 $\sqrt{\rho}$ \\
        $5$ &   0.786 &   28.125 &   0.182$\rho^{1/3}$ &   0.184$\rho^{1/3}$ \\
        $6$ &   0.848 &   35.39 &   0.254$\rho^{1/4}$ &   0.257$\rho^{1/4}$ \\
    [0.5ex] \hline
    \end{tabular}%
    \caption{Critical temperature $T_{c}$ for $b=0$ and
        $m^{2}=-2$. Here we choose the step size $\Delta
        b=0.1$.}\label{t1}
\end{table}
Therefore, the re-scaled critical temperature increases with
increasing the dimension for fixed values of the mass of the
scalar field and small values of the nonlinear parameter $b$. From
these tables we also understand that for each $d$, the critical
temperature decrease as the nonlinear parameter $b$ increases for
the fixed scalar field mass. So the condensation gets harder as
the nonlinear parameter becomes larger. This result is consistent with the earlier findings \cite%
{SS,SS2,SS3,ZPCJ}. Fig.(\ref{fig2}) represents a comparison between these
results from numeric and analytic calculations with different
values of the step size $\Delta b$.
\section{Condensation values and critical exponent} \label{Exp}
\subsection{Analytical method}
In this subsection we will calculate the order parameter $\langle\mathcal{O_{+}}%
\rangle$ as well as the critical exponent in the boundary of
spacetime. For this purpose we need the behaviour of the gauge
field $\phi$ near the critical point. If we write down Eq.(\ref{phiz}) near the critical point $T_{c}$ and keep terms that
are linear in $b$, we obtain

\begin{eqnarray}\label{pphi}
\phi^{\prime\prime}(z)+\frac{(4-d)}{z}\phi^{\prime }(z)+\frac{(-2+d)z^3
b\phi^{\prime 3}(z)}{r_{+}^{2}}-2\phi(z) r_{+}^{2}\frac{\psi^2(z)}{z^4 f(z)}%
\bigg(1-\frac{3 b z^4 \phi^{\prime 2}(z) }{2 r_{+}^{2}}\bigg)=0.
\end{eqnarray}
In the previous section we calculate the solution for this
equation in the case that we are at the critical point ($\psi=0$),
that obtained in the form Eq.(\ref{phi0}). Now in this section we
consider that the temperature is near the critical temperature, so
we have condensation and $\psi\neq0$, thus we use Eq.(\ref{psiF})
for $\psi$. Since we are near to critical point, the condensation
value is very small, and we can expand the solution for Eq.(\ref{pphi})
around the solution for $\phi$ at $T=T_{c}$ (that we had previously obtained it as Eq.(\ref{phi0})),
in terms of small parameter $\frac{\langle\mathcal{O_{+}}%
\rangle^{}2}{r_{+}^{2\Delta_{+}}}$, as
\begin{eqnarray}\label{bastphi}
\frac{\phi(z)}{r_+}=\lambda\zeta(z)+\frac{\langle\mathcal{O_{+}}\rangle^2}{%
r_+^{2\Delta_{+}}}\chi(z)+\cdot\cdot\cdot~,
\end{eqnarray}%
where we have taken the boundary condition as $\chi (1)=\chi
^{^{\prime }}(1)=0$. Substituting Eq.(\ref{psiF}) and
(\ref{bastphi}) into (\ref{pphi}), we arrive at the equation for
$\chi$
\begin{eqnarray}\label{chichi}
\chi^{\prime \prime }(z)+\left[(3d-6)b\lambda^{2} z^{3} \zeta^{\prime 2}(z)+%
\frac{4-d}{z}\right]\chi^{\prime }(z)=\frac{2z^{-4+2\Delta_{+}}
r_{+}^{2}\lambda
F^{2}(z)\zeta(z)}{f(z)}\left(1-\frac{3b\lambda^{2}z^{4}\zeta ^{\prime 2}(z)}{%
2}\right).
\end{eqnarray}
The left hand side of this equation can be rewritten as
\begin{eqnarray}\label{chichi1}
&&\chi''(z)+\left[(3d-6)b\lambda^{2}z^{3}\zeta'^{2}(z)+\frac{4-d}{z}\right]\chi'(z)=\chi''(z)+(3d-12)b\lambda^{2}z^{3}\zeta'^{2}(z)\chi'(z)
\nonumber\\
&&+6b\lambda^{2} z^{3}\zeta'^{2}(z)\chi'(z)
+\frac{4-d}{z}\chi'(z).
\end{eqnarray}
Taking into account the fact that
\begin{eqnarray}\label{aa}
3b\lambda^{2}z^{4}\zeta^{\prime }(z)\zeta^{\prime \prime }(z)
\chi^{\prime }(z)=3b\lambda^{2}z^{4}\zeta^{\prime }(z)
\frac{\phi^{\prime \prime}(z)}{\lambda r_{+}} \chi^{\prime }(z),
\end{eqnarray}
 if we rewrite Eq.(\ref{pphii}),
we find
\begin{eqnarray}
\phi^{\prime\prime}+\frac{bz^3d{\phi^{\prime}}^{3}}{2r_{+}^{2}}\left(1+\frac{bz^4}{2r_{+}^{2}}{\phi^{\prime}}^{2}\right)^{-1}+
\frac{(4-d)\phi^{\prime}}{z}\left(1+\frac{bz^4}{2r_{+}^{2}}{\phi^{\prime}}^{2}\right)^{-1}=0.
\end{eqnarray}
and substituting $\phi^{\prime \prime}(z)$ from this into (\ref{aa}), We have,
\begin{eqnarray}
&&3b\lambda^{2}z^{4}\zeta^{\prime }(z)\zeta^{\prime \prime }(z)
\chi^{\prime }(z)=3b\lambda^{2}z^{4} \frac{\zeta^{\prime
}(z)}{\lambda r_{+}}
\bigg(\frac{d-4}{z}\phi^{\prime}+O(b)+O(b^2) \bigg)\chi^{\prime }(z)\\ \nonumber
&&=3(d-4)b\lambda^2 z^3 \zeta^{\prime}
\frac{\phi^{\prime}}{\lambda r_{+}} \chi^{\prime
}(z)+O(b^2)+O(b^3)=3(d-4)b\lambda^{2}z^{3}\zeta^{\prime2}(z) \chi^{\prime
}(z).
\end{eqnarray}
Therefore we have
\begin{eqnarray}
3(d-4)b\lambda^{2}z^{3}\zeta^{\prime
2}(z) \chi^{\prime }(z)=3b\lambda^{2}z^{4}\zeta^{\prime }(z)\zeta^{\prime \prime
}(z) \chi^{\prime }(z),
\end{eqnarray}
and hence Eq.(\ref{chichi1}) may be written
\begin{eqnarray}\label{chiequation}
&&\chi''(z)+3b\lambda^{2}z^{4}\zeta'(z)\zeta''(z) \chi'(z)+6b\lambda^{2} z^{3}\zeta'^{2}(z)\chi'(z) +\frac{4-d}{z}\chi'(z)\nonumber\\
&&=\frac{2z^{-4+2\Delta_{+}} r_{+}^{2}\lambda
F^{2}(z)\zeta(z)}{f(z)}\left(1-\frac{3b\lambda^{2}z^{4}\zeta'^{2}(z)}{2}\right).
\end{eqnarray}
Multiplying both sides of Eq.(\ref{chiequation}) by the following
factor,
\begin{eqnarray}
T(z)=\frac{1}{z^{d-4}}e^{\frac{3b\lambda^{2}z^{4}\zeta^{\prime
2}(z)}{2}},
\end{eqnarray}
we can write Eq.(\ref{chiequation}) as,
\begin{eqnarray}
 (T\chi')^{\prime}=\frac{2z^{-d+2\Delta_{+}} r_{+}^{2}\lambda F^{2}(z)\zeta(z)}{f(z)}.
\end{eqnarray}
Integrating the above
equation in the interval $[0,1]$ and using the boundary conditions for $\chi$, yields
\begin{eqnarray}
T(z) \chi^{\prime
}(z)|_{z\rightarrow0}=-\int_{0}^{1}\frac{2z^{-d+2\Delta_{+}}
r_{+}^{2}\lambda F^{2}(z)\zeta(z)}{f(z)}dz.
\end{eqnarray}
Substituting $T(z)$ in above equation and noting that we have two
cases for $\zeta(z)$ to substitute in this equation, we finally
obtain
\begin{eqnarray}\label{chi-A}
\left[\frac{\chi^{\prime }(z)}{z^{d-4}}\right]\bigg|_{z\rightarrow
0}=-\lambda \mathcal{A},
\end{eqnarray}
\begin{eqnarray}
\mathcal{A}=\left\{
\begin{array}{rl}
\mathcal{A}_{1}~~~~~~ & \quad \mathrm{for}\
2(d-3)^{2}b_{n}(\lambda^{2}|_{b_{n-1}})<1, \\
&  \\
(\mathcal{A}_{2}+\mathcal{A}_{3}) \ , & \quad \mathrm{for}\
2(d-3)^{2}b_{n}(\lambda^{2}|_{b_{n-1}})>1,%
\end{array}%
\right.
\end{eqnarray}
where
\begin{eqnarray}
\mathcal{A}_{1}=\int_{0}^{1}\frac{2 r_{+}^2
z^{2\Delta_{+}-d}F^{2}\zeta_{1}}{f}dz,~~
\mathcal{A}_{2}=\int_{0}^{z_{0}}\frac{2 r_{+}^2 z^{2\Delta_{+}-d}F^{2}\zeta_{2}}{f}%
dz,~ \mathcal{A}_{3}=\int_{z_{0}}^{1}\frac{2 r_{+}^2 z^{2\Delta_{+}-d}F^{2}\zeta_{3}}{%
f}dz,
\end{eqnarray}
where $\zeta_{1}$, $\zeta_{2}$ and $\zeta_{3}$ are given by Eqs.(\ref{zeta1}), (\ref{zeta2}) and (\ref{zeta3}). Now we write down
the relation between $\chi^{\prime }(z)$ and $(d-3)$-th derivative
of $\chi(z)$. If we rewrite Eq.(\ref{chichi}) at $z\rightarrow0$,
we have
\begin{eqnarray}
\chi^{\prime \prime }(0)=\chi^{\prime }\left(\frac{d-4}{z}\right)\big|%
_{z\rightarrow0},
\end{eqnarray}
thus is a matter of calculations to show that we can write the
following relation in $d$-dimensions,
\begin{eqnarray}
\frac{\chi^{(d-3)}(z=0)}{(d-4)!}=\frac{\chi^{\prime }}{z^{d-4}}\big|%
_{z\rightarrow0}.
\end{eqnarray}
From Eqs.(\ref{bound1}) and (\ref{bastphi}) and by expanding
$\chi(z)$ around $z=0$, we have
\begin{eqnarray}
\mu-\frac{\rho}{r_{+}^{d-3}} z^{d-3}=r_{+} \lambda
\zeta+\frac{\langle
\mathcal{O_{+}}\rangle^2}{r_{+}^{2\Delta_{+}-1}}~
\Bigg\{\chi(0)+z\chi^{\prime}(0)+...+z^{( d-3)}
\frac{\chi^{(d-3)}(0)}{(d-3)!}+...\Bigg\}.
\end{eqnarray}
Comparing the coefficient of $z^{d-3}$ on both sides of the above
equation we obtain
\begin{eqnarray}\label{order-chi}
-\frac{\rho}{r_{+}^{d-2}}=-\lambda +\frac{\langle\mathcal{O_{+}}\rangle^2}{%
r_{+}^{2\Delta_{+}}}~ \frac{\chi^{d-3}(0)}{(d-3)!}~,
\end{eqnarray}
Using Eq.(\ref{order-chi}) and (\ref{chi-A}), we arrive at
\begin{eqnarray}
\frac{\rho}{r_{+}^{d-2}}=\lambda \left[1+\frac{\langle\mathcal{O_{+}}\rangle^2}{%
r_{+}^{2\Delta_{+}}} \frac{\mathcal{A}}{(d-3)}\right],
\end{eqnarray}
with regards to definition of $\lambda=\frac{\rho}{r_{+c}^{d-2}}$,
and substituting $r_{+}$ and $r_{+c}$ from the relations that we
have for $T$ and $T_{c}$ given in Eqs.(\ref{T}) and (\ref{TTc}),
we find the relation between the condensation operator and the
critical temperature in $d$-dimensional spacetime near the
critical temperature ($T\sim T_{c}$) as
\begin{eqnarray}
\langle\mathcal{O_{+}}\rangle=\left(\frac{4\pi}{d-1}\right)^{\Delta_{+}} \sqrt{\frac{%
(d-3)(d-2)}{\mathcal{A}}}~T_{c}^{\Delta_{+}}
\sqrt{1-\frac{T}{T_{c}}}.
\end{eqnarray}
Thus, we find that the critical exponent of the order parameter is
$1/2$, and near the critical point this operator satisfies
\begin{eqnarray}\label{orderparameter}
\langle\mathcal{O_{+}}\rangle=\beta T_{c}^{\Delta_{+}}\left(1-\frac{T}{T_c}\right)^{%
{1}/{2}},
\end{eqnarray}
which holds for various values of $b$, $m$ and $d$. The
coefficient $\beta$ is given by
\begin{eqnarray}\label{beta}
\beta=\left\{
\begin{array}{rl}
\left(\frac{4\pi}{d-1}\right)^{\Delta_{+}}\sqrt{\frac{(d-3)(d-2)}{\mathcal{A}_{1}}}\
,~~~~
& \quad \mathrm{for}\ 2(d-3)^{2} b_{n}(\lambda^{2}|_{b_{n-1}})<1, \\
& \\
\left(\frac{4\pi}{d-1}\right)^{\Delta_{+}}\sqrt{\frac{(d-3)(d-2)}{\mathcal{A}_{2}+\mathcal{A}_{3}}}%
\ , & \quad \mathrm{for}\ 2(d-3)^{2} b_{n}(\lambda^{2}|_{b_{n-1}})>1.%
\end{array}%
\right.
\end{eqnarray}
From these results we can analysis the effect of the nonlinear
parameter $b$ and the spacetime dimension $d$, on the values of
$\beta$. Our analytical results are presented in table (\ref{t4}),(\ref{t5}) and (\ref{t6})
which we also compare them with the numerical results.
\subsection{Numerical method}
In the previous section for the numerical solution, we was needed
only the charge density at the critical point for obtaining the
re-scaled critical temperature. Here we start with increasing $\psi(1)$ from $\psi(1)=\frac{1}{%
10000}$ to higher values in the small steps, meaning that the
temperature becomes lower. At any step we can find all the
coefficient of the asymptotic behavior of $\psi$ and $\phi$, such
as $\psi_{+}$. We use the
value of $\psi_{+}$ for calculation of the order parameter $\langle\mathcal{O_{+}%
}\rangle$, and exploring the behavior of this parameter in terms
of temperature for different dimension of the spacetime and for
different values
of $b$. For example for $d=4,5,6$ we obtain condensation $\langle\mathcal{O_{+}}%
\rangle$ from following relations,
\begin{eqnarray}
\langle\mathcal{O_{+}}\rangle&=&\sqrt{2}\psi_{+}\quad\mathrm{for}~~ d=4, \\
\langle\mathcal{O_{+}}\rangle&=&\psi_{+}\quad ~~~~\mathrm{for}~~~
d=5,6,
\end{eqnarray}
where the coefficient $\sqrt{2}$ is a convenient normalization
factor \cite{Har1}. Now we want to plot the dimensionless
condensation as a function of dimensionless temperature. Since we
work in units where $c=\hbar=1$, all physical quantities can be
described in unit which is some power of the mass. In this unit,
length and time have dimension of $ [mass]^{-1}$, energy, momentum
and $T$ have dimension $[mass]$, while $\rho$ has dimension $ [mass]^{d-2}$. Also since in this unit, the scalar field must be
dimensionless, so $\psi_{+}$  must have dimension
$[mass]^{\Delta_{+}}$.\\
Thus we can plot dimensionless
${\langle\mathcal{O_{+}}\rangle}/{T_{c}^{\Delta_{+}}}$ as a
function of ${T}/{T_{c}}$, where $\Delta_{+}$ is defined by Eq.(\ref{delta}).
\begin{figure*}[h]
\begin{center}
\begin{minipage}[b]{0.325\textwidth}\begin{center}
        \subfigure[~$d=4$]{
                 \label{fig3a}\includegraphics[width=\textwidth]{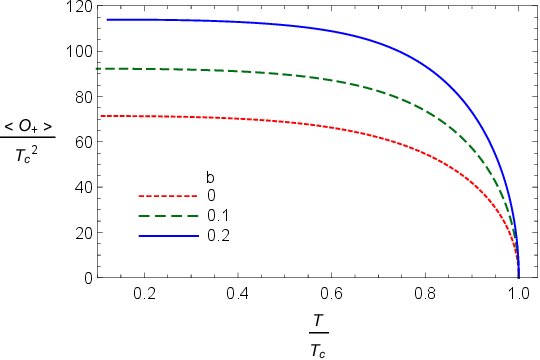}\qquad}
     \end{center}\end{minipage} \hskip+0cm
\begin{minipage}[b]{0.325\textwidth}\begin{center}
         \subfigure[~$d=5$]{
                  \label{fig3b}\includegraphics[width=\textwidth]{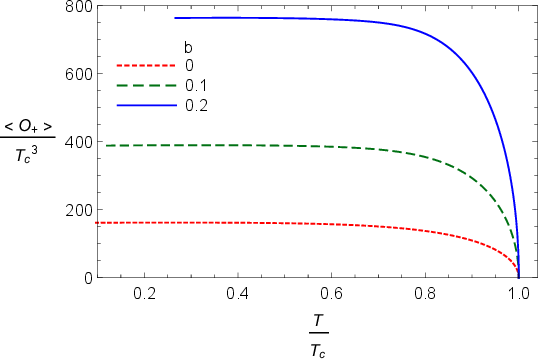}\qquad}
     \end{center}\end{minipage} \hskip0cm
\begin{minipage}[b]{0.325\textwidth}\begin{center}
          \subfigure[~$d=6$]{
                   \label{fig3c}\includegraphics[width=\textwidth]{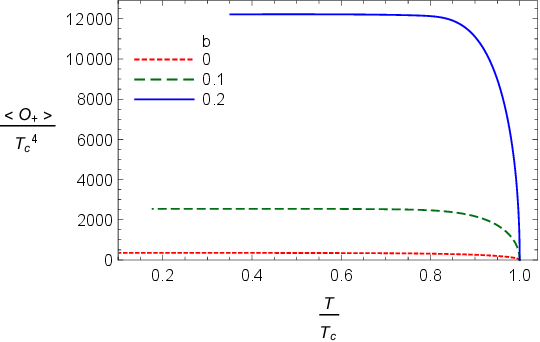}\qquad}
     \end{center}\end{minipage} \hskip0cm
\end{center}
\caption{The dimensionless condensation operator in terms of dimension less
temperature for different values of $b$.}
\label{fig3}
\end{figure*}
\begin{figure}[H]
\centering
\subfigure[$b=0$] {\includegraphics[scale=0.85]{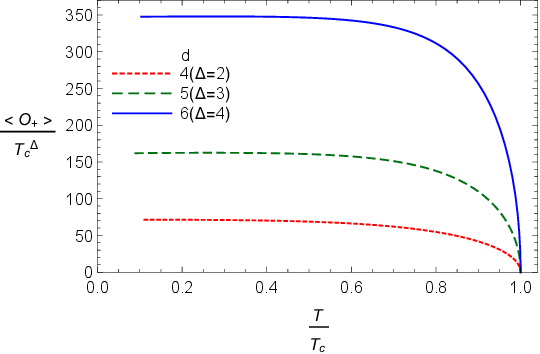}} \label{fig4a}
\hspace*{.1cm} \subfigure[$b=0.1$] {\includegraphics[scale=0.85]{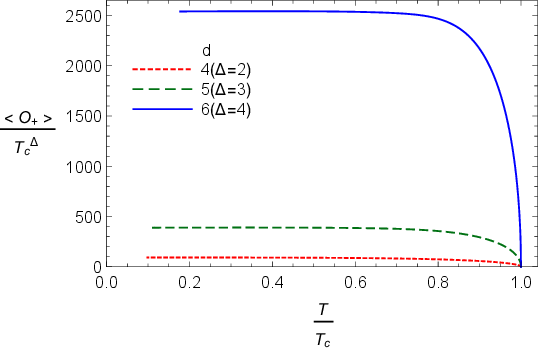}} %
\label{fig4b}
\caption{ The dimensionless condensation operator in terms of dimensionless
temperature for different values of $d$.}
\label{fig4}
\end{figure}
This curves for the condensation operator are qualitatively
similar to what that obtained in BCS theory, the condensate rises
quickly when the system is stayed on below the critical temperature and goes to a constant as $%
T\rightarrow0$. Near the critical temperature, as obtained from analytical
results in Eq.(\ref{orderparameter}), the condensate is proportional to $(1-\frac{T}{T_{c}}%
)^{1/2}$, that is the behavior that predicted by Landau-Ginzburg
theory. This curves for $d=4,5,6$ in Fig.(\ref{fig3}) represents
that, when we increase $b$, the dimensionless condensation becomes larger. Also,
by comparing the condensation for different $d$ in Fig.(\ref{fig4}), we find that it becomes large in higher dimensions.

\begin{figure*}[h]
\begin{center}
\begin{minipage}[b]{0.325\textwidth}\begin{center}
         \subfigure[~$d=4$]{
                  \label{fig5a}\includegraphics[width=\textwidth]{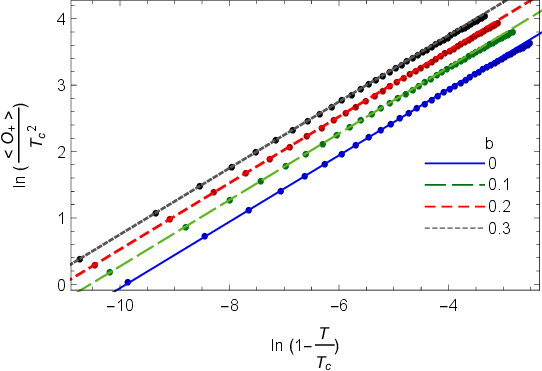}\qquad}
      \end{center}\end{minipage} \hskip+0cm
\begin{minipage}[b]{0.325\textwidth}\begin{center}
          \subfigure[~$d=5$]{
                   \label{fig5b}\includegraphics[width=\textwidth]{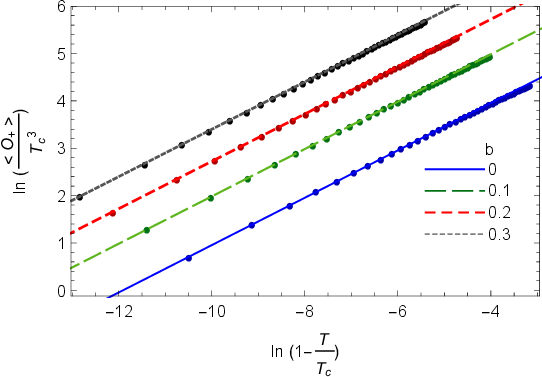}\qquad}
      \end{center}\end{minipage} \hskip0cm
\begin{minipage}[b]{0.325\textwidth}\begin{center}
           \subfigure[~$d=6$]{
                    \label{fig5c}\includegraphics[width=\textwidth]{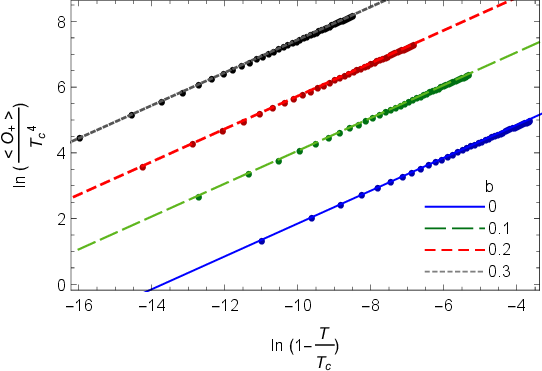}\qquad}
      \end{center}\end{minipage} \hskip0cm
\end{center}
\caption{Fitting the order parameter with a straight line whose slope is $%
{1}/{2}$ for different $b$ and $d$.} \label{fig5}
\end{figure*}

Now we find that the results obtained for the behaviour of the
condensation operator near the critical point, from numerical
calculation is in good agreement with the results obtained from
analytical calculation in Eq.(\ref{orderparameter}). From Eq.(\ref{orderparameter}) we can write
\begin{eqnarray}
\ln
\left(\frac{\langle\mathcal{O_{+}}\rangle}{T_{c}^{\Delta_{+}}}\right)=\ln
\beta +\frac{1}{2}\ln \left(1-\frac{T}{T_{c}}\right),
\end{eqnarray}
Now we can plot $\ln \left(\frac{\langle\mathcal{O_{+}}\rangle}{T_{c}^{\Delta_{+}}}%
\right)$ as a function of $\ln(1-\frac{T}{T_{c}})$. From the
dotted curves in Fig.(\ref{fig5}), we see that the plot which is
fitted to a straight line has slop $1/2$, that is the critical
exponent. The slope is independent of parameters $b$ and $d$. Also
we can find $\beta$ from the y-intercept of the lines. Finally we
conclude that the phase transition is of second order and the
critical exponent of the system always take the value $1/2$, and
the nonlinear electrodynamics can not change the result. This
result seems to be a universal property for various nonlinear
electrodynamics \cite{GR,SS3,SSM}.

Now we summarize the results for $\beta$ for a holographic
superconductor in logarithmic electrodynamics,
which obtained from analytical calculation from Eq.(\ref{beta})
and from numerical calculation which explained before, for
different values of $b$ and $d$, in tables (\ref{t4}),(\ref{t5}) and (\ref{t6}). Also we compare
these results.
\begin{table}[H]
\centering
\begin{tabular}{cccccccc}
\hline\hline b &  & $\gamma |_{\rm Analytical}$ &  & $\gamma
|_{\rm Numerical}$   \\ \hline
0 &  & 92.8 &  & 140.09  \\
0.1 &  & 108.88 &  & 194.52   \\
0.2 &  & 124.03 &  & 251.09   \\
0.3 &  & 282.83 &  & 314.32  \\ \hline

\end{tabular}%
\caption{The values of
$\protect\gamma=\protect\sqrt{2}\protect\beta$ for $d=4$ and the
step size $\Delta b=0.1$. }\label{t4}
\end{table}
The results that obtained for $\beta$ from the numerical and
analytical solutions for $d=5$ and $d=6$ represent in the last
tables. Also for these dimensions, the results from analytical for
the step size $\Delta b=0.1$ is far from the numerical, so we
consider a smaller step size $\Delta b=0.05$, to increase our
accuracy.
\begin{table}[H]
\centering
\par
\begin{tabular}{cccccccccc}
\hline\hline b &  & $\beta |_{\rm Analytical(1)}$ & &$\beta |_{\rm
Analytical(2)}$ & & $\beta |_{\rm Numerical}$   \\ \hline
0 &  & 238.5 &  & 238.5 & & 385.20  \\
0.1 &  & 330.93 &  & 369.70 & & 1077.9  \\
0.2 &  & 441.06 &  & 537.81 & & 2259.4 \\
0.3 &  & 580.26 &  & 779.34 & & 4445.8 \\ \hline
\end{tabular}%
\caption{The values of $\protect\beta$ for $d=5$ and the step size
for Analytical(1) is $\Delta b=0.1$ and for Analytical(2) $\Delta
b=0.05$.}\label{t5}
\end{table}

\begin{table}[H]
\centering
\par
\begin{tabular}{ccccccccc}
\hline\hline b &  & $\beta |_{\rm Analytical(1)}$ & &$\beta |_{\rm
Analytical(2)}$ & & $\beta |_{\rm Numerical}$   \\ \hline
0 &  & 533.82 &  & 533.82 & & 934.97  \\
0.1 &  & 871.26 &  & 1088.12 & & 8629.5  \\
0.2 &  & 1350.20 &  & 2920.86 & & 45677 \\
0.3 &  & 2071.40 &  & 8213.86 & & 253670 \\ \hline
\end{tabular}%
\caption{The values of $\protect\beta$ for $d=6$ and the step size
for Analytical (1) is $\Delta b=0.1$ and for Analytical(2) $\Delta
b=0.05$.}\label{t6}
\end{table}
From these tables we find that the value $\beta$,
increases with increasing the nonlinear parameter $b$. Also, when
the step size $\Delta b$ is smaller, the analytical results are
closer with the numerical results rather than the larger step
size.
\section{Conductivity} \label{Cond}
\begin{figure*}[h!]
\begin{center}
\begin{minipage}[b]{0.325\textwidth}\begin{center}
                    \subfigure[ ~$d=4,~b=0$]{
                        \label{fig6a}\includegraphics[width=\textwidth]{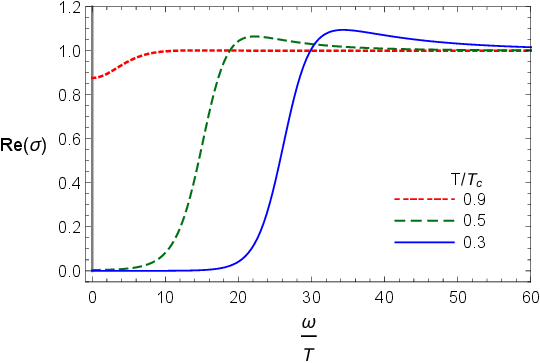}\qquad}
            \end{center}\end{minipage} \hskip+0cm
\begin{minipage}[b]{0.325\textwidth}\begin{center}
                    \subfigure[ ~$d=4, ~b=0.1$]{
                        \label{fig6b}\includegraphics[width=\textwidth]{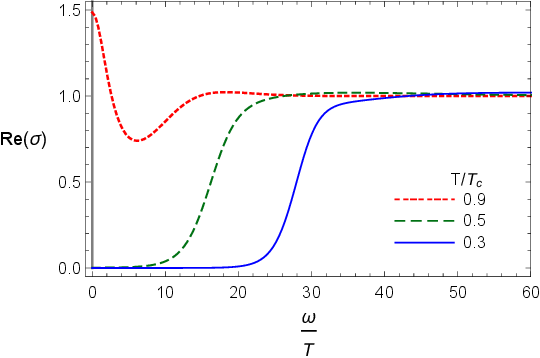}\qquad}
            \end{center}\end{minipage} \hskip0cm
\begin{minipage}[b]{0.325\textwidth}\begin{center}
                    \subfigure[ ~$d=5,~b=0$]{
                        \label{fig6c}\includegraphics[width=\textwidth]{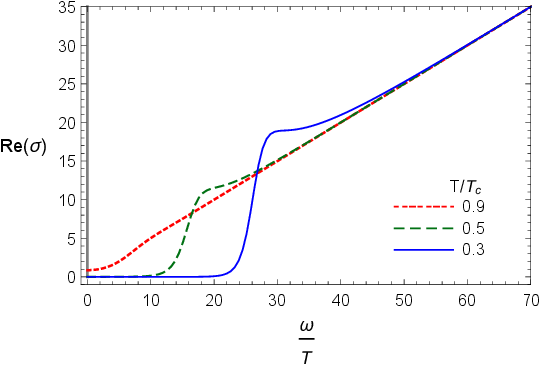}\qquad}
            \end{center}\end{minipage} \hskip0cm
\begin{minipage}[b]{0.325\textwidth}\begin{center}
                    \subfigure[$~ d=5,~b=0.1$]{
                        \label{fig6d}\includegraphics[width=\textwidth]{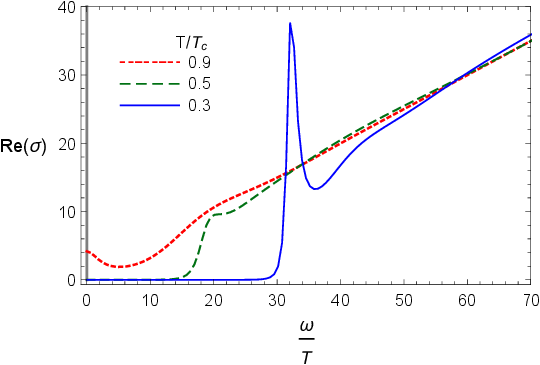}\qquad}
            \end{center}\end{minipage} \hskip0cm
\begin{minipage}[b]{0.325\textwidth}\begin{center}
                    \subfigure[$~ d=6,~b=0$]{
                        \label{fig6e}\includegraphics[width=\textwidth]{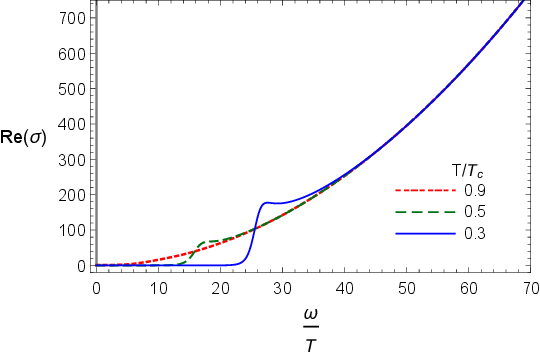}\qquad}
            \end{center}\end{minipage} \hskip0cm
\begin{minipage}[b]{0.325\textwidth}\begin{center}
                    \subfigure[$~ d=6,~b=0.1$]{
                        \label{fig6f}\includegraphics[width=\textwidth]{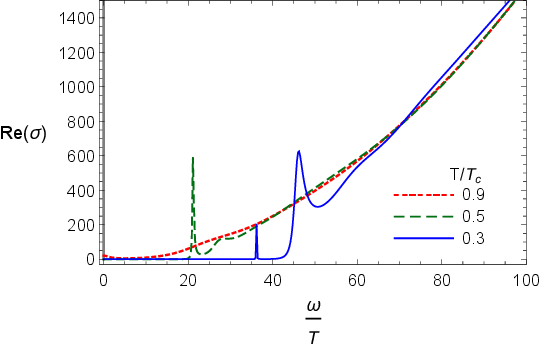}\qquad}
            \end{center}\end{minipage} \hskip+0cm
\end{center}
\caption{The real part of conductivity for different temperature in terms of
$\protect\omega/T$. }
\label{fig6}
\end{figure*}

\begin{figure}[H]
\begin{center}
\begin{minipage}[b]{0.325\textwidth}\begin{center}
                    \subfigure[ ~$d=4,~b=0$]{
                        \label{fig7a}\includegraphics[width=\textwidth]{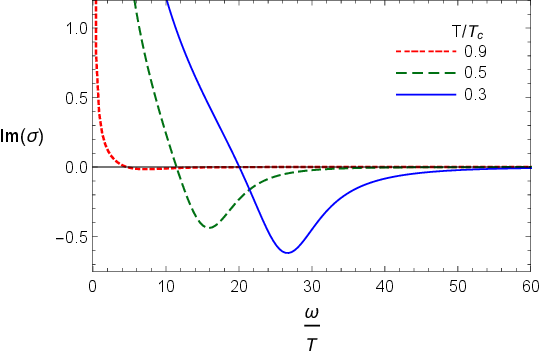}\qquad}
            \end{center}\end{minipage} \hskip+0cm
\begin{minipage}[b]{0.325\textwidth}\begin{center}
                    \subfigure[ ~$d=4, ~b=0.1$]{
                        \label{fig7b}\includegraphics[width=\textwidth]{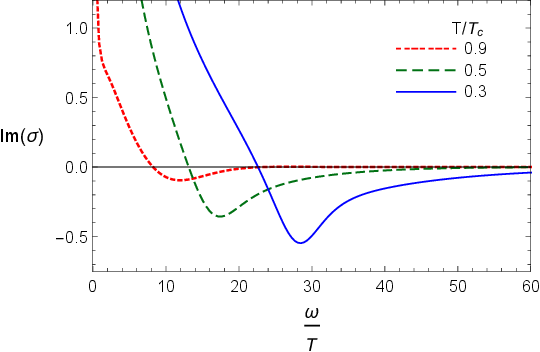}\qquad}
            \end{center}\end{minipage} \hskip0cm
\begin{minipage}[b]{0.325\textwidth}\begin{center}
                    \subfigure[ ~$d=5,~b=0$]{
                        \label{fig7c}\includegraphics[width=\textwidth]{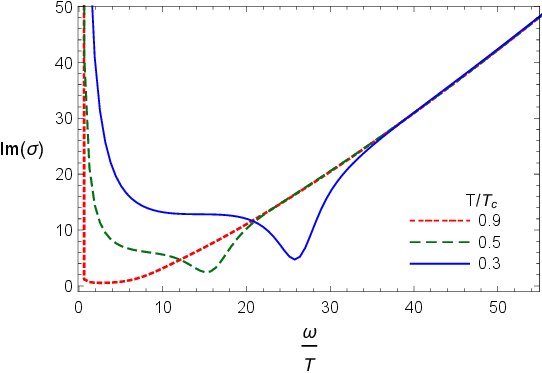}\qquad}
            \end{center}\end{minipage} \hskip0cm
\begin{minipage}[b]{0.325\textwidth}\begin{center}
                    \subfigure[$~ d=5,~b=0.1$]{
                        \label{fig7d}\includegraphics[width=\textwidth]{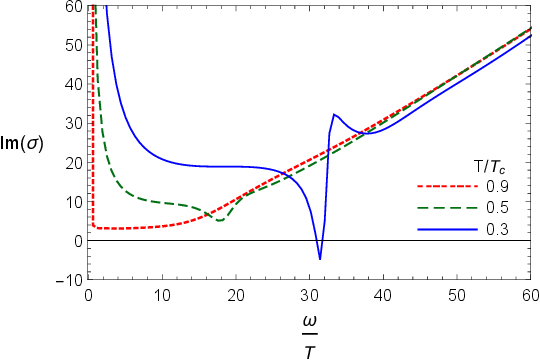}\qquad}
            \end{center}\end{minipage} \hskip0cm
        \begin{minipage}[b]{0.325\textwidth}\begin{center}
                \subfigure[$~ d=6,~b=0$]{
                    \label{fig7e}\includegraphics[width=\textwidth]{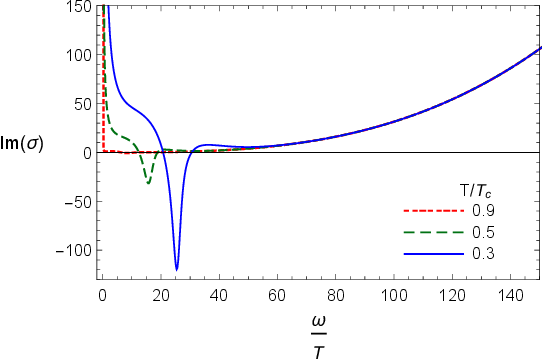}\qquad}
        \end{center}\end{minipage} \hskip0cm
    \begin{minipage}[b]{0.325\textwidth}\begin{center}
            \subfigure[$~ d=6,~b=0.1$]{
                \label{fig7f}\includegraphics[width=\textwidth]{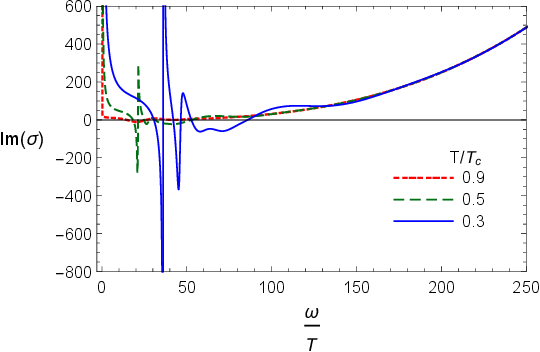}\qquad}
    \end{center}\end{minipage} \hskip0cm
\end{center}
\caption{The imaginary part of conductivity for different temperature in
terms of $\protect\omega /T$. }
\label{fig7}
\end{figure}

\begin{figure*}[h]
\begin{center}
\begin{minipage}[b]{0.325\textwidth}\begin{center}
                \subfigure[$~ d=4,~T/T_{c}\approx0.15$]{
                    \label{fig8a}\includegraphics[width=\textwidth]{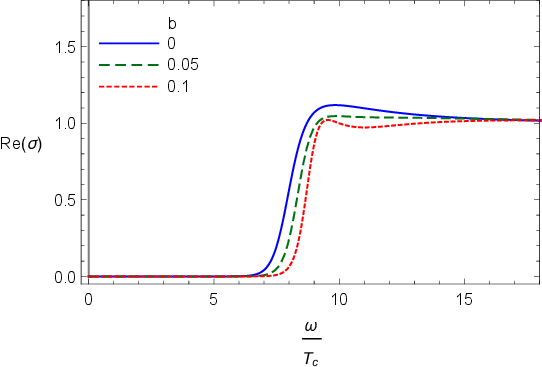}\qquad}
        \end{center}\end{minipage} \hskip+0cm
\begin{minipage}[b]{0.325\textwidth}\begin{center}
                \subfigure[$~ d=5,~T/T_{c}\approx0.15$]{
                    \label{fig8b}\includegraphics[width=\textwidth]{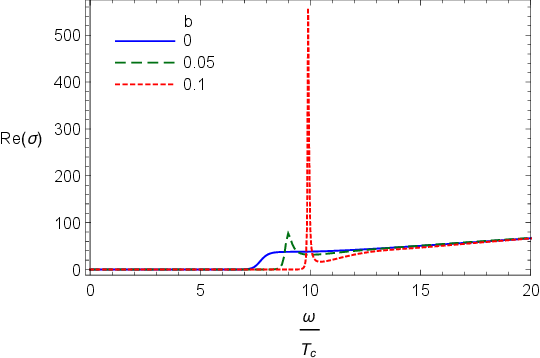}\qquad}
        \end{center}\end{minipage} \hskip0cm
\begin{minipage}[b]{0.325\textwidth}\begin{center}
                \subfigure[$~ d=6,~T/T_{c}\approx0.15$]{
                    \label{fig8c}\includegraphics[width=\textwidth]{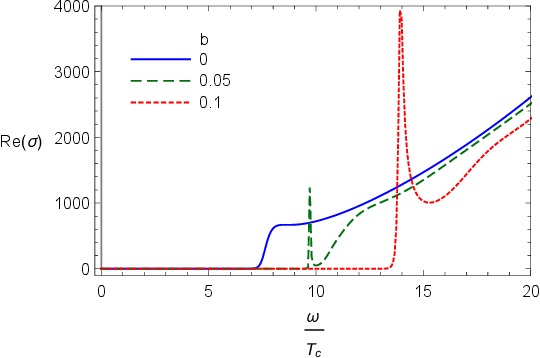}\qquad}
        \end{center}\end{minipage} \hskip0cm
\end{center}
\caption{The real part of conductivity for different $b$ in terms of $%
\protect\omega /T_{c}$.}
\label{fig8}
\end{figure*}

\begin{figure}[H]
    \begin{center}
        \begin{minipage}[b]{0.325\textwidth}\begin{center}
                \subfigure[$~ d=4$]{
                    \label{fig9a}\includegraphics[width=\textwidth]{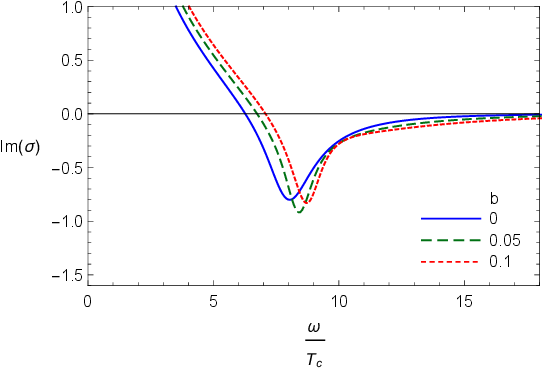}\qquad}
        \end{center}\end{minipage} \hskip+0cm
        \begin{minipage}[b]{0.325\textwidth}\begin{center}
                \subfigure[$~ d=5$]{
                    \label{fig9b}\includegraphics[width=\textwidth]{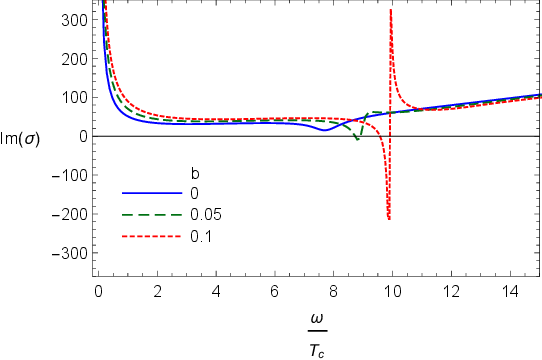}\qquad}
        \end{center}\end{minipage} \hskip0cm
        \begin{minipage}[b]{0.325\textwidth}\begin{center}
                \subfigure[$~ d=6$]{
                    \label{fig9c}\includegraphics[width=\textwidth]{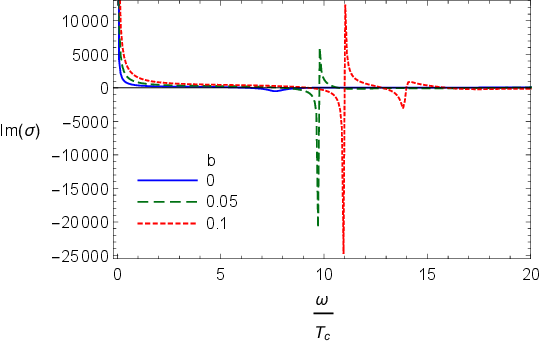}\qquad}
        \end{center}\end{minipage} \hskip0cm
    \end{center}
    \caption{The real part of conductivity for different $b$ in terms of $%
        \protect\omega /T_{c}$.}
    \label{fig9}
\end{figure}

The superconductor energy gap is an essential feature of the
superconducting state which may be characterized by the threshold
frequency obtained from the electrical conductivity. Hence, in
this section we investigate the behavior of the electric
conductivity as a function of frequency. In the linear response
theory, the conductivity is expressed as the current density
response to an applied electric field
\begin{equation}
\sigma _{ij}=\frac{J_{i}}{E_{j}}~.  \label{con}
\end{equation}%
According to the AdS/CFT correspondence dictionary, if we want to
have current in the boundary, we must consider a vector potential
in the bulk. This implies that by solving for fluctuations of the
vector potential $A_{j}$
in the bulk, we will have a dual current operator $J_{i}$ in the CFT \cite%
{Har1}. Inasmuch as the dual CFT has a spatial symmetry, one can
consider just the conductivity in the $x$ direction. We turn on
the small perturbation in the bulk gauge potential as
\begin{equation*}
\delta A_{x}=A_{x}(r)e^{-i\omega t}~,
\end{equation*}%
where $\omega $ is the frequency. Thus, the equation of motion for
$A_{x}(r)$, at the linearized level of the perturbation, takes the
form
\begin{eqnarray}
&&\big(2-b\phi ^{\prime 2}(r)\big)A_{x}^{\prime \prime }(r)+\bigg[2b\phi ^{\prime }(r)\phi
^{\prime \prime }(r)+\left( \frac{d-4}{r}+\frac{f^{\prime }(r)}{f(r)}\right)
\big(2-b\phi ^{\prime 2}(r)\big)\bigg]A_{x}^{\prime }(r)  \notag \\
&&+\frac{\omega ^{2}}{f^{2}(r)}\big(2-b\phi ^{\prime 2}(r)\big)A_{x}(r)-\frac{\psi ^{2}(r)}{%
f(r)}(2-b\phi ^{\prime 2}(r))^{2}A_{x}(r)=0.   \label{Ax}
\end{eqnarray}%
The asymptotic ($r\rightarrow \infty $) behavior of the above differential
equation is obtained as
\begin{equation*}
A_{x}^{\prime \prime }(r)+\frac{d-2}{r} A_{x}^{\prime
}(r)+\frac{\omega^{2}}{r^4} A_{x}(r)=0,
\end{equation*}
which admits the following solution in the asymptotic $(r\rightarrow\infty)$,
\begin{equation}\label{aaa}
A_{x}=\left\{
\begin{array}{rl}
A^{(0)}+\frac{A^{(1)}}{r}-\frac{\omega ^{2}A^{(0)}}{2r^{2}}...\ , & \ \ ~~%
\mathrm{for}\ d=4, \\
& \\
A^{(0)}+\frac{A^{(1)}}{r^{2}}+\frac{\omega ^{2}A^{(0)}ln(kr)}{2r^{2}}+..., &
\ \quad \mathrm{for}\ d=5, \\
& \\
A^{(0)}+\frac{A^{(1)}}{r^3}+\frac{\omega ^{2}A^{(0)}}{2r^{2}}...\ , & \ \ ~~%
\mathrm{for}\ d=6, \\
& \\
\end{array}%
\right.
\end{equation}%
where $A^{(0)},A^{(1)}$ are constant parameters and $k$ is a
constant with $[\textit{length}]^{-1}$ dimension which inserted
for a dimensionless logarithmic argument. From the AdS/CFT
dictionary, the boundary current operator may be
calculated by differentiating the action \cite{D.Tong}%
\begin{equation}
J=\frac{\delta S_{bulk}}{\delta A^{(0)}}=\frac{\delta S_{o.s}}{\delta A^{(0)}}=\frac{\partial \left( \sqrt{-g}%
\mathcal{L}_{m}\right) }{\partial A_{x}^{^{\prime }}}~\big|_{r\rightarrow
\infty },  \label{current}
\end{equation}%
where $A^{(0)}$ is the dual to a source in the boundary theory.
Also, $S_{o.s}$ and $\mathcal{L}_{m}$ are, respectively, the
on-shell bulk action and the Lagrangian of the
matter field. The $S_{o.s}$ action is given by%
\begin{equation}
S_{o.s}=\int_{r_{+}}^{\infty }dr\int
d^{d-1}x\sqrt{-g}\mathcal{L}_{m}.
\end{equation}%
Expanding the action to quadratic order in the perturbation and
taking into account Eq.(\ref{Ax}), $S_{o.s.}$ reduces to%
\begin{equation}
S_{o.s}=\int d^{d-1}x~\frac{r^{d-4}f(r)A_{x}(r)A_{x}^{\prime }(r)}{-2+b\phi
^{\prime 2}(r)}\big|_{r\rightarrow \infty }.
\end{equation}%
According to the asymptotic behavior of $\phi $ and $A_{x}$ given by Eq.(\ref%
{bound1}) and Eq.(\ref{aaa}), and using Eq.(\ref{current}%
), one can calculate the holographic current as%
\begin{equation}
J_{x}=\left\{
\begin{array}{rl}
A^{(1)}\ , & \ \ ~~\mathrm{for}\ d=4 \\
&  \\
2A^{(1)}-\frac{\omega ^{2}A^{(0)}}{2}, & \ \quad \mathrm{for}\
d=5, \\
&  \\
3A^{(1)}\ , & \ \ ~~\mathrm{for}\ d=6. \\
&  \\
\end{array}%
\right.
\end{equation}%
Thus, from Eq. (\ref{con}) and $E_{x}=-\partial _{t}\delta A_{x}$,
the electrical conductivity is obtained as%
\begin{equation}
\sigma =\left\{
\begin{array}{rl}
\frac{A^{(1)}}{i\omega A^{(0)}}\ , & \ \ ~~\mathrm{for}\ d=4,\\
&  \\
\frac{2A^{(1)}}{i\omega A^{(0)}}+\frac{i\omega }{2}, & \ \quad
\mathrm{for}\ d=5,\\
&  \\
\frac{3A^{(1)}}{i\omega A^{(0)}}\ , & \ \ ~~\mathrm{for}\ d=6.\\
&  \\
\end{array}%
\right.  \label{conduc}
\end{equation}%
It is worth noting that the divergence terms in the above action
is eliminated by adding a suitable counterterm
\cite{skenderis},\cite{Ruth}. Now, one can numerically solve the
differential equation for $A_{x}(r)$ in Eq.(\ref{Ax}) by imposing
an ingoing wave boundary condition near the event horizon
\cite{SAH}
\begin{equation}
A_{x}(r)=S(r)f^{{-i\omega }/{(4\pi T)}},
\end{equation}%
where
\begin{equation*}
S(r)=1+a_{1}(r-r_{+})+a_{2}(r-r_{+})^{2}+...,
\end{equation*}%
$T$ is the Hawking temperature and the coefficients $a_{1}$, $a_{2}$, $%
\ldots $ are characterized by Taylor expansion of Eq.(\ref{Ax})
around the horizon $r_{+}$. With $A_{x}$ at hand, we can calculate
the conductivity from Eq.(\ref{conduc}). We summarize our results
regarding the behaviour of the conductivity in Figs.(\ref{fig6}-\ref{fig9}).

The behavior of the real parts of conductivity as a function of
frequency for various nonlinear parameter $b$ and in various
dimension at different temperature are depicted in
Fig.(\ref{fig6}). As one can see from this figure, the
superconducting gap appears below the critical temperature that
becomes deep with decreasing the temperature.  That means
$\omega_{g}$ becomes larger. Since $\omega_{g}$ is probational to
the minimum energy that needed to break the condensation,
therefore with decreasing the temperature, the condensation
becomes stronger. Also, the gap becomes sharper as we decrease the
temperature. At enough large frequency, the behavior of
conductivity indicates a normal state that follows a power law
relation with frequency, i.e. $\mathrm{Re}[\sigma ]\varpropto
\omega ^{d-4}$ \cite{D.Tong}. For $3$-dimension of CFT, the real
part of conductivity is independent of frequency which tends
toward a constant value for large frequency (see Figs.\ref{fig6a}
and \ref{fig6b}).

The associated imaginary parts of conductivity are illustrated in
Fig. \ref{fig7} which is related to the real parts of conductivity
by the Kramers--Kronig relations. Hence, the pole in the imaginary
parts of conductivity at $\omega =0$ points out to a delta
function in the real parts which are shown by the vertical lines
in Fig. \ref{fig6}. Although, the delta function cannot be
resolved numerically, but we know that it exists. By comparison
the figures, we find that at any fixed temperature and frequency,
the conductivity in higher dimensions is larger. For $d=6$, more
delta functions and poles appear inside the gap as one decreases
the temperature. The BCS theory explains systems that are weak
coupled, which means, there was no interaction between the pairs.
But holographic superconductors are strongly coupled. With
decreasing the temperature, the interactions become stronger and
form a bound state. The additional delta functions and poles
related to this state \cite{Gary T.H}.

In order to determine the effect of the dimension and nonlinear
parameter on the superconducting gap at low temperature $T\approx
0.15T_{c}$, we plot real and imaginary parts of the holographic
electrical conductivity as a function of normalized frequency
$\omega /T_{c}$ in Figs.\ref{fig8} and \ref{fig9}. From the BCS
theory we have relation $\omega_{g}=2 \Delta$, where $\Delta$ is
the energy required for charged excitations, that leads to
$\omega_{g}\simeq 3.5 T_{c}$. In \cite{Gary T.H}, it was shown
that the relation connecting the frequency gap  with the critical
temperature, for $d=3$ and $d=4$ dimensional holographic
superconductor becomes ${\omega_{g}}/{T_{c}}\approx8$, which is
more than twice of the corresponding value in the BCS theory. Also
it was argued that this ratio for $d=4,5$, is always about eight,
and the relation $\omega_{g}/T_{c}\approx8$ is universal. However,
as one can see from Fig.\ref{fig8}  in each dimension, the
superconducting gap increases with increasing the nonlinear
parameter $b$. Also, for the fixed value of the nonlinear
parameter $b$, the energy gap effectively increases with
increasing the dimension, which indicates that the holographic
superconductor state is destroyed for large $\omega /T_{c}$. This
implies that the relation between $\omega_{g}$ and $T_{c}$ depends
on the parameters $b$ and $d$.

\section{Closing Remarks}
To sum up, in this paper we have continued the study on the
gauge/gravity duality by investigating the properties of the
$s$-wave holographic superconductor in higher dimensional
spacetime and in the presence of nonlinear gauge field. We have
considered the Logarithmic Lagrangian for the $U(1)$ gauge theory
which was proposed by Soleng \cite{Soleng}. We follow the
Sturm-Liouville eigenvalue problem for our analytical study as
well as the numerical shooting method. We explored three aspects
of these kinds of superconductors. First, we obtained the relation
between critical temperature and charge density, $\rho$, and
disclosed the effects of both nonlinear parameter $b$ and the
dimensions of spacetime, $d$, on the critical temperature $T_c$.
We found that in each dimension, $T_c/{\rho}^{1/(d-2)}$ decreases
with increasing the nonlinear parameter $b$. Besides, for a fixed
value of $b$, this ratio increases for the higher dimensional
spacetime. This implies that the high temperature superconductor
can be achieved in the higher dimensional spacetime. We confirmed
that our analytical method is in good agreement with the numerical
results. Second, we have calculated the condensation value and
critical exponent of the system analytically as well as
numerically and observed that in each dimension, the coefficient
$\beta$ becomes larger with increasing the nonlinear parameter
$b$. Besides, for a fixed value of $b$, it increases with
increasing the spacetime dimension, i.e., in higher dimensional
spacetime.

Finally, we explored the electrical conductivity of the
holographic superconductor. Our aim in this part was to disclose
the effects of the nonlinear gauge field as well as the higher
dimensional spacetime on the superconducting gap  of the
holographic superconductor. We observed that the superconducting
gap appears below the critical temperature that becomes deep with
decreasing the temperature. Besides, we found that at high
frequency, the behavior of conductivity indicates a normal state
that follows a power law relation with frequency, i.e.
$\mathrm{Re}[\sigma ]\varpropto \omega ^{d-4}$. We also
investigated the imaginary part of superconductor and found that
the pole in the imaginary parts of conductivity at $\omega =0$
points out to a delta function in the real parts. We concluded
that for a fixed value of the nonlinear parameter $b$, the energy
of gap effectively increases with increasing the dimension, which
indicates that the holographic superconductor state is destroyed
for large $\omega /T_{c}$. This indicates that the relation
between $\omega_{g}$ and $T_{c}$ depends on the parameters $b$ and
$d$.

\section*{Acknowledgments} We are grateful to the referee for
constructive comments which helped us improve our paper
significantly. We thank Shiraz University Research Council. The
work of A.S has been supported financially by Research Institute
for Astronomy and Astrophysics of Maragha (RIAAM), Iran.

\end{document}